\documentclass[sigconf, 10pt]{acmart}
\usepackage{subfigure}
\usepackage{pifont}
\usepackage[ruled,linesnumbered, vlined]{algorithm2e}
\usepackage{multirow}
\usepackage{enumitem}
\usepackage{colortbl}
\usepackage{tabularx}
\usepackage{bbding}

\usepackage{pifont}
\usepackage{tikz}
\newcommand*\circled[1]{\tikz[baseline=(char.base)]{
            \node[shape=circle,draw,white, fill=black,inner sep=1pt] (char) {#1};}}
\newcommand*\squared[1]{\tikz[baseline=(char.base)]{
            \node[shape=rectangle,draw,white, fill=black, minimum size=10pt, inner sep=1pt] (char) {#1};}}

\newcommand{\highlight}[1]{\textcolor{black}{#1}}
\newcommand{\revise}[1]{\textcolor{black}{#1}}
\AtBeginDocument{%
  \providecommand\BibTeX{{%
    \normalfont B\kern-0.5em{\scshape i\kern-0.25em b}\kern-0.8em\TeX}}}

\copyrightyear{2024}
\acmYear{2024}
\setcopyright{acmlicensed}\acmConference[ACM MobiCom '24]{The 30th Annual International Conference on Mobile Computing and Networking}{September 30-October 4, 2024}{Washington D.C., DC, USA}
\acmBooktitle{The 30th Annual International Conference on Mobile Computing and Networking (ACM MobiCom '24), September 30-October 4, 2024, Washington D.C., DC, USA}
\acmDOI{10.1145/3636534.3649363}
\acmISBN{979-8-4007-0489-5/24/09}

\begin{document}

\title[Asteroid]{Asteroid: Resource-Efficient Hybrid Pipeline Parallelism for Collaborative DNN Training on Heterogeneous Edge Devices}



\author{Shengyuan Ye$^\blacklozenge$$^*$, Liekang Zeng$^\blacklozenge$$^*$, Xiaowen Chu$^\blacktriangle$, Guoliang Xing$^\lozenge$, Xu Chen$^\blacklozenge$$^\dagger$}
\affiliation{
  \country{$^\blacklozenge$School of Computer Science and Engineering, Sun Yat-sen University, Guangzhou, China}
}
\affiliation{
\country{$^\blacktriangle$Data Science and Analytics Thrust, HKUST (GZ), Guangzhou, China}
}
\affiliation{
\country{$^\lozenge$The Chinese University of Hong Kong, Hong Kong SAR, China}
}
\email{{yeshy8,zenglk3}@mail2.sysu.edu.cn, xwchu@ust.hk, glxing@cuhk.edu.hk, chenxu35@mail.sysu.edu.cn}
\thanks{$*$: Equal contributions. $\dagger$: Corresponding author.}






\renewcommand{\shortauthors}{Shengyuan Ye, et al.}

\begin{abstract}
On-device Deep Neural Network (DNN) training has been recognized as crucial for privacy-preserving machine learning at the edge. However, the intensive training workload and limited onboard computing resources pose significant challenges to the availability and efficiency of model training.
While existing works address these challenges through native resource management optimization, we instead leverage our observation that 
edge environments usually comprise a rich set of accompanying trusted edge devices with idle resources beyond a single terminal.
We propose Asteroid, a distributed edge training system that breaks the resource walls across heterogeneous edge devices for efficient model training acceleration.
\revise{Asteroid adopts a hybrid pipeline parallelism to orchestrate distributed training}, along with a judicious parallelism planning for maximizing throughput under certain resource constraints.
Furthermore, a fault-tolerant yet lightweight pipeline replay mechanism is developed to tame the device-level dynamics for training robustness and performance stability.
We implement Asteroid on heterogeneous edge devices with both vision and language models, demonstrating up to $12.2\times$ faster training than conventional parallelism methods and $2.1\times$ faster than state-of-the-art hybrid parallelism methods through evaluations.
Furthermore, Asteroid can recover training pipeline $14\times$ faster than baseline methods while preserving comparable throughput despite unexpected device exiting and failure.
\end{abstract}


\maketitle

\section{Introduction}
\label{sec:intro}

Deep Neural Networks (DNNs) have driven diverse intelligence in today's smart applications, ranging from voice assistance, smart robotics to city surveillance, etc.
While existing works have extensively studied the inference aspect of DNN models, the growing proliferation of human-in-the-loop intelligent services urgently emphasizes the necessity for privacy-preserving personalization and continuous model refinement, raising the need for advanced on-device learning ability.
For instance, in Federated Learning \cite{mcmahan2017communication, cai2022autofednlp}, user devices are required to provision a local model training task in order to contribute to and share the learning procedure.
In Continual Learning \cite{bhardwaj2022ekya, xu2018deeptype}, user devices periodically retrain their local models with newly-collected data so as to adapt the model performance to the contextual factors.

Despite the increasing demand, efficient in-situ learning still suffers from its prohibitively long training time and vulnerable convergence stability.
As we will empirically show in \S\ref{sec:training_on_device}, even training a mobile-oriented compact DNN model in typical edge devices (i.e., Jetson Nano) takes 160$\times$ longer epoch time than that in a GPU server, showing the fundamental contradiction between intensive training workload and constrained on-board resources. Moreover, insufficient memory capacity can be a real game-stopper for on-device learning.
Towards alleviating these issues, existing wisdom has explored extensive optimizations from various aspects.
For example, a number of works adopt model compression techniques (e.g., pruning, sparsification, and quantization) or manually designed lightweight model architectures to reduce the computation complexity for DNN training \cite{jiang2022model, xu2019elfish, sandler2018mobilenetv2, bhattacharya2016sparsification, lin2022device}.
Other leading research works have explored to design sophisticated management mechanisms (e.g., tensor rematerialization \cite{patil2022poet, jain2020checkmate, chen2016training}, memory budget adapting \cite{wang2022melon, gim2022memory}) on native resources, but are still bottlenecked by the intrinsic deficiency of physical resource shortage.

In this paper, we alternatively observe that prevalent edge scenarios like smart homes and smart factories usually comprise a group of trusted idle devices beyond a single terminal (e.g., pads, laptops, and smart-home devices owned by the same user or family) \cite{ye2022eco, zeng2020coedge, zhao2018deepthings}.
These accompanying devices are typically in physical proximity to the primary one running on-device learning tasks and can be associated as a resource augmentation for in-situ DNN training acceleration.
This motivates us to regard adjacent available devices as a resource pool and collaborate with them in a distributed manner to render expedited model training at the edge.

We note that distributed training acceleration has been comprehensively studied in cloud datacenters for years, but is much less understood with edge devices.
Nevertheless, scheduling efficient distributed training over edge devices is non-trivial given the unique challenges inherent in edge environments: 
(1) In contrast to accelerator clusters in cloud, edge devices are extremely limited in terms of computing power, memory capacity, and communication bandwidth.
(2) Edge devices are much more heterogeneous compared to cloud server configurations, which necessitates a heterogeneity-aware strategy to maximize the utilization of computing potential.
(3) Edge devices frequently exhibit more potential dynamics than dedicated cloud environments, due to the devices' mobility and accessibility.
Unfortunately, no existing work can address all of the aforementioned challenges.

To this end, we propose \textbf{Asteroid}, a general distributed training system that is able to orchestrate multiple heterogeneous edge devices for expedited, resource-efficient, and fault-tolerant model training.
Asteroid's contribution goes beyond merely leveraging distributed edge devices for training acceleration, instead it addresses the above challenges in three levels.
\revise{First, from a parallelism perspective, a hybrid pipeline parallelism (HPP) is employed as a principle to manage the distributed training workflow, which combines the best of data parallelism and pipeline parallelism and allows significantly larger optimization space for parallelism planning in heterogeneous edge environments}. 
Second, to maximize resource utilization of HPP among heterogeneous edge devices, a novel dynamic-programming based parallelism planning algorithm is designed, as well as a memory-efficient batch ingestion strategy for multidimensional resources optimization including memory budget, limited communication capacity, and resource heterogeneity.
Finally, to adapt to dynamic participants during the runtime, a fault-tolerant mechanism is applied through lightweight coarse-granularity workload migration and topology-driven model replication. 
We implement Asteroid in four realistic testbeds with each consisting of at least 4 heterogeneous edge devices.
Extensive evaluations on both vision and language models show that Asteroid is able to deliver up to 12.2$\times$ speedup over conventional parallel training counterparts,
and achieves up to 2.1$\times$ throughput improvement over the state-of-the-art hybrid parallelism methods.
Besides, Asteroid can well adapt to device-level dynamics (i.e., device exiting or failure) in agile pipeline recovering time (14$\times$ faster than baseline) with minimal throughput sacrifice.

The main contributions are summarized as follows.

\vspace{-0.2cm}

\begin{itemize}[leftmargin=*]
    \item Through extensive measurement studies on on-device and parallel training performance, \revise{we employ a hybrid pipeline parallelism as a principled tool to collaborate trusted edge devices for model training acceleration}.
    \item We design a novel planning algorithm tailored for hybrid parallelism mechanism, which comprehensively considers memory budget, limited communication capacity, and resource heterogeneity of edge devices.
    \item We propose Asteroid, a general distributed edge training system that is able to orchestrate resource-efficient, expedited training across heterogeneous edge devices with fault-tolerant robustness. Asteroid reveals another path towards efficient in-situ model training at the edge.
    \item We implement Asteroid and evaluate it in realistic heterogeneous testbeds. Experimental results show up to 12.2$\times$ throughput improvement over baselines and strong robustness to device dynamics with swift pipeline recovering.
\end{itemize}

\vspace{-0.5cm}

\section{Motivation and Preliminaries}
\label{background}
\subsection{DNN Training on Resource-Constrained Edge Devices}
\label{sec:training_on_device}
On-device training can leverage locally collected data to improve model performance while fully preserving data in-situ, making it a widely utilized approach in privacy-sensitive edge applications \cite{zhou2019edge, mcmahan2017communication, bhardwaj2022ekya, ouyang2021clusterfl, shuai2022balancefl}. However, the resource-intensive and computation-demanding nature of DNN training presents significant challenges for resource-constrained edge devices \cite{xu2022mandheling, lin2022device,wang2022melon, gim2022memory}.

To gain insights, we conduct several experiments to analyze how limited computation resource affects on-device DNN training. 
Specifically, we perform on-device learning for three widely-used DNN architectures on off-the-shelve edge devices and Nvidia GPU platform for comparison, as shown in Table \ref{tab:average_epoch_time}. 
We observe that the average time of a training epoch exhibits a huge gap between A100 and Jetson boards, e.g., 160$\times$ and 67$\times$ slowdown for Jetson Nano and TX2 when comparing with A100 on MobileNetV2 \cite{sandler2018mobilenetv2}, a representative compact DNN architecture specific for mobile platforms.
Memory capacity is another crucial knob of DNN training. However, contemporary edge platforms are incapable of accommodating the memory demands of prominent state-of-the-art DNN models despite their latest release \cite{wang2022melon}.

\begin{table}[t]
\small
\caption{Elapsed time of a training epoch on devices.}
\vspace{-7pt}
\begin{tabular}{cccc}
\hline
\multirow{2}{*}{\begin{tabular}[c]{@{}c@{}}DNN Model\end{tabular}} & \multicolumn{3}{c}{Average Epoch Time} \\ \cline{2-4} 
                                                                     & A100    & Jetson TX2   & Jetson Nano   \\ \hline 
EfficientNet-B1                                                      & 10sec     & 11.2min      & 26.7min       \\
MobileNetV2                                                          & 9.4sec    & 8.5min       & 22min         \\ 
ResNet50                                                             & 65sec     & 1.14hour     & 3.48hour      \\ \hline
\end{tabular}
\label{tab:average_epoch_time}
\vspace{-0.6cm}
\end{table}

To mitigate resource constraints, existing works \cite{xu2019elfish, jiang2019model, matsubara2020head, chen2021quantization, chen2020knowledge} use compression techniques like pruning, quantization, and knowledge distillation to build crafted models dedicated to edge platforms. While these techniques reduce computation, they compromise model accuracy. Other works, such as compilation optimization \cite{gim2022memory} or run-time memory management \cite{wang2022melon, patil2022poet}, can better utilize device resources, but are still restricted by the physical resources of a single device.

Alternatively, in typical edge scenarios such as smart homes, there are usually multiple trusted edge devices in physical proximity that are owned by the same user or family, such that mutually-trustworthy computation resource sharing among these can be achieved. Some pioneering research works have dived into collaborative edge computing to break the resource walls across edge devices \cite{ye2024galaxy, wei2024communication, zeng2020coedge, zhao2018deepthings, zeng2022fograph}. However, most works focus only on model inference and few of them manage to address model training.

\vspace{-0.1cm}

\subsection{\revise{Edge Collaborative Training with Data Parallelism and Pipeline Parallelism}}

\begin{figure}[t]
    \setlength{\abovecaptionskip}{0.1cm}
    \setlength{\belowcaptionskip}{-0.4cm}
    \centering
    \includegraphics[width=0.95\linewidth]{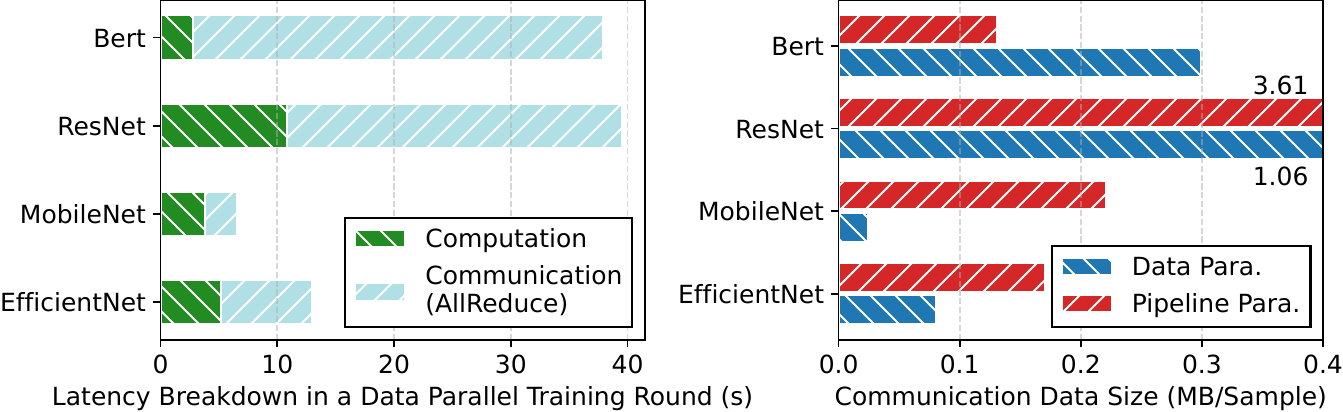}
    \caption{\highlight{Left: The training latency breakdown in DP. Right: Bytes communicated per sample in DP and PP. Both experiments are conducted on a three-Jetson Nano edge environment with 100Mbps D2D bandwidth.}}
    \label{fig:motivation_dp}
\end{figure}

\textbf{Data Parallelism.} The most common way to train DNN models in parallel is \textit{data parallelism} (\textbf{DP}) \cite{li2014communication, goyal2017accurate}. In DP, inputs are partitioned across workers, and each worker maintains a replica of the entire model and performs training on its local data while periodically synchronizing gradients with other workers (i.e., AllReduce). 
The simplicity of its workload induces better scalability with multiple devices.
However, due to the loose and varying edge connections, the communication overhead caused by synchronization can usually dominate training time, as shown in Fig. \ref{fig:motivation_dp}(Left).

\textbf{Pipeline Parallelism.} Another widely-used parallelism is \textit{pipeline parallelism} (\textbf{PP}). PP is an advanced model parallelism-based training strategy that executes the DNN model in a pipelined manner across multiple workers  \cite{huang2019gpipe}. Specifically, in PP, the DNN model is partitioned into multiple stages and each stage is mapped to a separate processor for stage-wise forward/backward pass execution. The partitioning ensures each stage has an approximately equal workload, optimizing parallel efficiency.
For language model (e.g., Bert-small \cite{devlin2018bert}) or crafted edge model with tiny activations, PP is far more communication-efficient than DP, since each worker only needs to exchange a subset of output activations with neighboring workers. 
Nonetheless, pipeline parallelism is also followed with shortcomings: (1) \textit{Weak scalability}. The straight-forward implementation of PP for edge clusters can create too many stages, which amplifies the impact of inter-stage communication latency. (2) \textit{Unoverlappable inter-stage communication}. When communication occurs between layers with huge intermediate activations, PP can not effectively overlap the inter-stage communication with forward and backward execution. \highlight{Our experiment, implemented with Gpipe \cite{huang2019gpipe} for PP, shows that the inter-stage communication latency can be up to $24\times$ longer than the stage execution time, which dominates the entire training process.
As shown in Fig. \ref{fig:motivation_dp}(Right), for CNN-based models, the per-sample data size communicated in PP even surpasses the communication volume necessitated by DP.}

\subsection{\revise{Key Insight: Combining Data Parallelism with Pipeline Parallelism}}
\begin{figure}[t!]
    \setlength{\abovecaptionskip}{0.1cm}
    \setlength{\belowcaptionskip}{-0.4cm}
    \centering
    \includegraphics[width=0.9\linewidth]{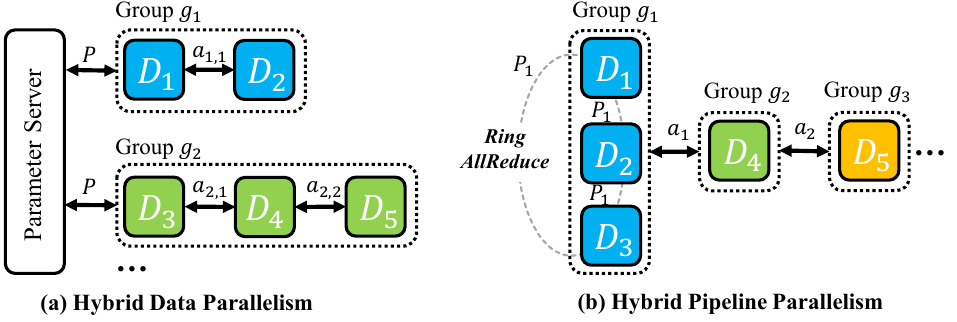}
    \caption{\revise{Illustration of HDP and HPP.}}
    \label{fig:HDP-HPP}
\end{figure}

\revise{The above analysis motivates us to employ a hybrid parallelism architecture that incorporates the best of both DP and PP, so as to achieve superior performance in complex and heterogeneous edge environments.
\textit{Hybrid Data Parallelism} (\textbf{HDP}), as adopted by Hetpipe \cite{park2020hetpipe}, organizes devices into groups for inter-group DP and intra-group PP, necessitating a centralized parameter server (PS) for full gradient exchange. 
Alternatively, another hybrid approach, \textit{Hybrid Pipeline Parallelism} (\textbf{HPP}), utilized by PipeDream \cite{narayanan2019pipedream} and Dapple \cite{fan2021dapple}, arranges devices into groups for inter-group PP and intra-group DP, as depicted in Fig. \ref{fig:HDP-HPP}.}

\vspace{-0.5cm}

\revise{To better understand the communication efficiency of these two architectures, we quantitatively analyze their communication volume across devices in a formal way.
We first focus on HDP.
Assume that each edge device can be and only be divided into a specific group $g_i$ and the considered architecture counts $G$ device groups in total.
Let $P$ be the gradient size of the global model, the communication volume for the parameter server bidirectional synchronization is $2GP$.
The pipelined communication data size within a group is $2\beta_i\sum_{j=1}^{|g_i|-1} a_{i, j}$, where $\beta_i$ is the batch size handled by group $g_i$ and $a_{i,j}$ is the size of $j$-th intermediate tensor in $g_i$.
When there are multiple device groups (e.g., in Fig. \ref{fig:HDP-HPP}(a)), we can summarize the above analysis to derive the case of $G>1$ in Eq. (\ref{eq:HDP}).
If there is exactly one device group ($G=1$), however, PS synchronization is free and thus only intra-group communication volume is charged. }
\vspace{-0.2cm}
\begin{align}
\mathcal{V}_{\text{HDP}} &= 
\begin{cases}
2GP + \sum_{i=1}^{G}\left(2\beta_i\sum_{j=1}^{|g_i|-1} a_{i, j}\right), & G>1, \\
2\beta_1\sum_{j=1}^{|g_1|-1} a_{1, j}, & G = 1.
\end{cases}
\label{eq:HDP}
\end{align}

\begin{figure*}[t]
    \setlength{\abovecaptionskip}{0.1cm}
    \setlength{\belowcaptionskip}{-0.4cm}
    \centering
    \includegraphics[width=0.9\linewidth]{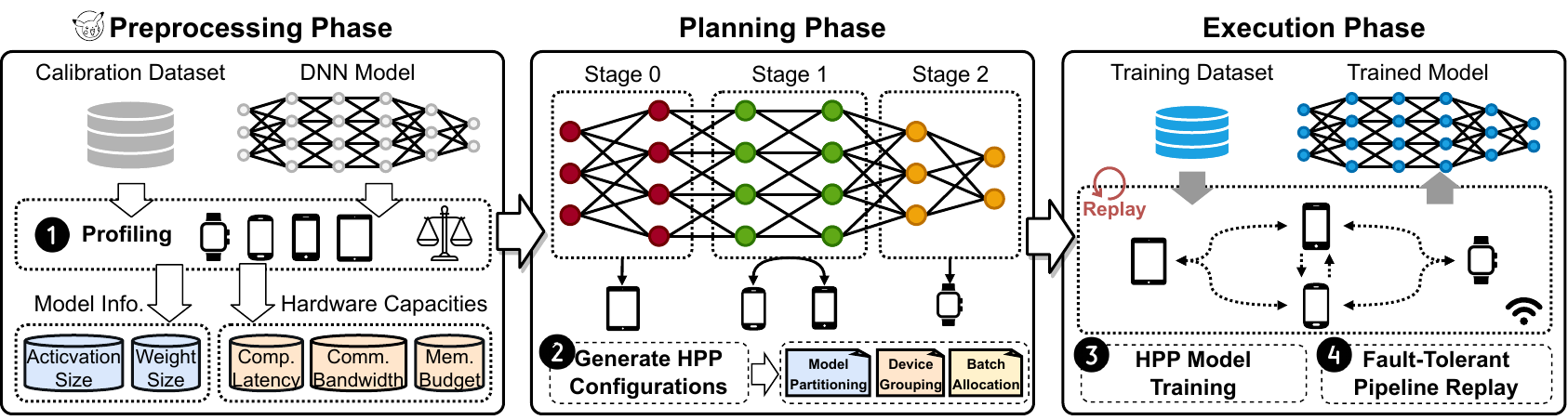}
    \caption{Asteroid Overview: A three-phase workflow includes Preprocessing, Planning, and Execution Phase.}
    \label{fig:overview}
\end{figure*}

\revise{We next target at HPP.
Different from HDP which shares the global model across groups, each group in HPP may take charge of only a part of the model, whose size is denoted as $P_i$ for group $g_i$.
For example, group $g_1$ in Fig. \ref{fig:HDP-HPP}(b) takes a model segmentation of size $P_1$.
Given $\beta$ as the global mini-batch size and $a_i$ as the size of the intermediate tensor exported by group $g_i$, each group in HPP requires $2(|g_i|-1)$ round of ring AllReduce of sub-model $P_i$ in group $g_i$ and the total intra-group communication is thus $\sum_{i=1}^{G}\left[2(|g_i|-1)P_i\right]$.
With the inter-group pipelined communication data size as $2\beta\sum_{j=1}^{G-1} a_j$, the total communication volume of HPP can be derived in the case of $G>1$ of Eq. (\ref{eq:HPP}).
When $G=1$, inter-group communications are eliminated and the formula can be simplified into the case of $G=1$ of Eq. (\ref{eq:HPP}).}
\vspace{-0.2cm}
\begin{align}
\mathcal{V}_{\text{HPP}} &= 
\begin{cases}
\sum_{i=1}^{G}\left[2(|g_i|-1)P_i\right] + 2\beta\sum_{j=1}^{G-1} a_j, & G>1, \\
2(|g_1|-1)P , & G = 1.
\end{cases}
\label{eq:HPP}
\end{align}

\revise{Upon the above formulas, we further empirically evaluate representative models in Table \ref{tab:average_epoch_time} in an edge environment with five Jetson Nano devices.  
For HDP, we adopted Hetpipe's allocation recommendations, and for HPP, we adopt Asteroid's planning method (refer to \S \ref{sec:algorithm}). The results in Table \ref{tab:comm_volume} reveal HDP's communication volume exceeds HPP's by $1.9\times$ - $2.7\times$. This is because HDP employed by HetPipe necessitates full parameter exchange between groups once the number of groups surpasses one. Conversely, HPP's architecture, through parallelism planning, confines AllReduce operations to initial parameter-light convolution layers, thereby circumventing the final parameter-dense layers.
}

\begin{table}[t!]
\small
\caption{\revise{Comparison of communication volume for training a global mini-batch employing HDP and HPP.}}
\vspace{-0.3cm}
\label{tab:comm_volume}
\begin{tabular}{cccc}
\hline
Volume & EfficientNet-B1 & MobileNetV2 & ResNet50 \\ \hline
$\mathcal{V}_{\text{HDP}}$ (MB) & 171.4 & 98.0 & 576.2 \\ 
$\mathcal{V}_{\text{HPP}}$ (MB) & 76.2 & 52.1 & 212.4 \\ \hline
\end{tabular}
\vspace{-0.7cm}
\end{table}

\revise{\textbf{Opportunities with HPP across Edge Devices.} 
In light of the foregoing analysis, Asteroid employs HPP to facilitate collaboration with edge devices.
Beyond breaking the resource wall of a single device, employing HPP offers the following benefits:
(1) Each device stores only a subset of the entire model, leading to a smaller memory footprint, which is particularly advantageous for models with huge parameters.
(2) HPP offers a highly flexible parallelism architecture that can effectively minimize communication volume by preventing AllReduce in parameter-dense layers.
(3) Through layer-wise planning, HPP can avoid inter-stage communication between layers with huge intermediate tensors. By fully overlapping computation and communication, HPP conceals the limitations of network capacity in edge environments.
(4) HPP provides higher scheduling flexibility and expands a larger optimization space of parallelism planning in our considered complex and heterogeneous edge environments.}

\vspace{-0.3cm}

\subsection{\revise{Technical Challenges}}
Despite the benefits, realizing HPP in complex edge environments still suffers from a set of challenges.

\textbf{Scarce Memory and Network Capacity.}
Edge devices, constrained by limited memory and bandwidth through links such as WiFi and cabled network \cite{huang2022real}, require meticulous planning for the HPP architecture to prevent out-of-memory issues and maximize bandwidth efficiency. Successfully leveraging the architecture's flexibility to address these constraints presents a significant challenge.

\textbf{Heterogeneous Computing Resource.} 
Edge environments are typically heterogeneous \cite{zhou2019edge}, with facilities ranging from small devices and gateways \cite{chen2018edge} to much powerful cloudlets \cite{satyanarayanan2009case}.
Efficiently applying HPP in such highly heterogeneous settings is a particularly challenging aim of maximizing its resource utilization, which requires judiciously matching workload distribution to diverse edge resources.

\textbf{Dynamic Training Participants.}
Further complicating the problem is the inherent, unpredictable dynamics of available edge resources, due to devices moving across networks and multitasking \cite{wang2022melon, venieris2022multi}. To render stable, reliable training performance with HPP therefore poses significant challenges in designing a robust scheduling such that training participants' failure is tolerable.


\begin{figure}[t]
    \setlength{\abovecaptionskip}{0.1cm}
    \setlength{\belowcaptionskip}{-0.5cm}
    \centering
    \subfigure[The computation graph is partitioned into three stages, where Stage 1 is replicated on a device group with two devices for intra-stage data parallelism.]{
        \begin{minipage}[t]{\linewidth}
        \centering
        \includegraphics[width=0.65\linewidth]{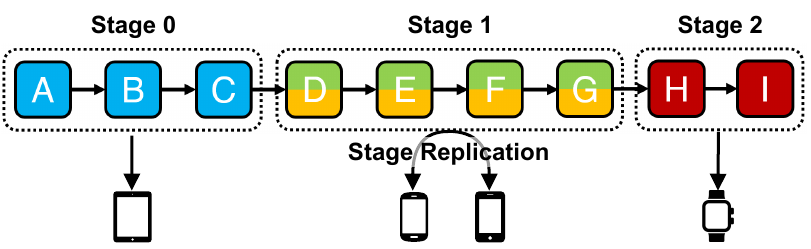}
        \label{fig:stage_partition}
        \end{minipage}
    }
    \subfigure[Training pipeline of 5 micro-batches. The numbers in the cells represent micro-batch ids. AllReduce is performed in Stage 1 for model synchronization. 
    The star marks the dominant step (will be introduced in \S\ref{sec:algorithm}).
    ]{
        \begin{minipage}[t]{\linewidth}
        \centering
        \includegraphics[width=0.95\linewidth]{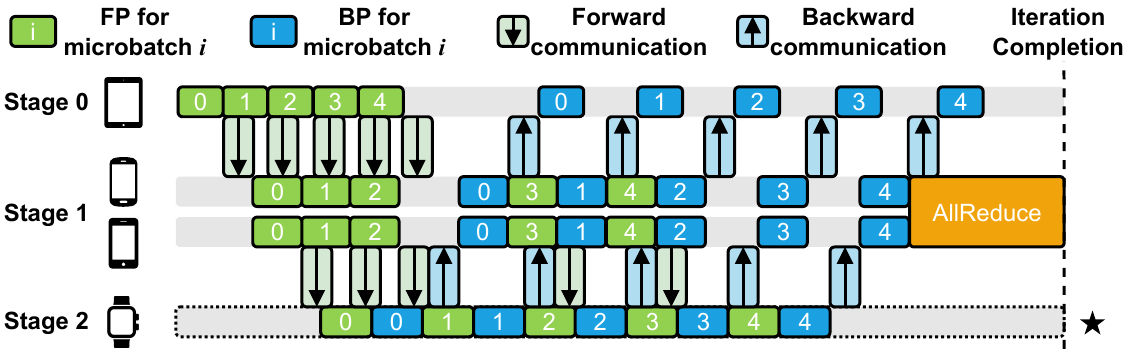}
        \label{fig:hpp_pipeline}
        \end{minipage}
    }
    \vspace{-0.2cm}
    \caption{An instance of HPP with four edge devices.}
    \label{fig:hybrid_pipeline_parallelism}
    \vspace{-0.2cm}
\end{figure}

\section{Asteroid System Design}
\label{sec:execution}
\subsection{System Overview}
Fig. \ref{fig:overview} depicts an overview of our proposed Asteroid which features three primary phases: \textit{Preprocessing Phase}, \textit{Planning Phase} and \textit{Execution Phase}. 
\textit{Preprocessing Phase} is an offline procedure that runs once before deployment. \textit{Asteroid Profiler} performs a training process using calibration data as input on the physical edge devices to record the runtime profile necessary for parallelism planning (step \circled{1}). 
\revise{During \textit{Planning Phase}, \textit{Asteroid Planner} takes profiling results as input to generate planning configurations that include DNN model partitioning points, device grouping strategies, and micro-batch allocations within groups (step \circled{2}).} These configurations comprehensively addresses challenges including memory budget, limited communication capacity, and resource heterogeneity, and is subsequently applied to target DNN models and training participants.
In \textit{Execution Phase}, \textit{Asteroid Worker} which is deployed on each participant will be responsible for model execution, intermediate output exchange and gradient synchronization (step \circled{3}). To account for the challenge of runtime dynamics, a fault-tolerant pipeline replay mechanism will be applied (step \circled{4}).
A central coordinator (a user-specified device) is required to apply planning configuration and detect device failure.

\begin{figure}[t!]
    \setlength{\abovecaptionskip}{0.1cm}
    \setlength{\belowcaptionskip}{-0.4cm}
    \centering
    \includegraphics[width=0.9\linewidth]{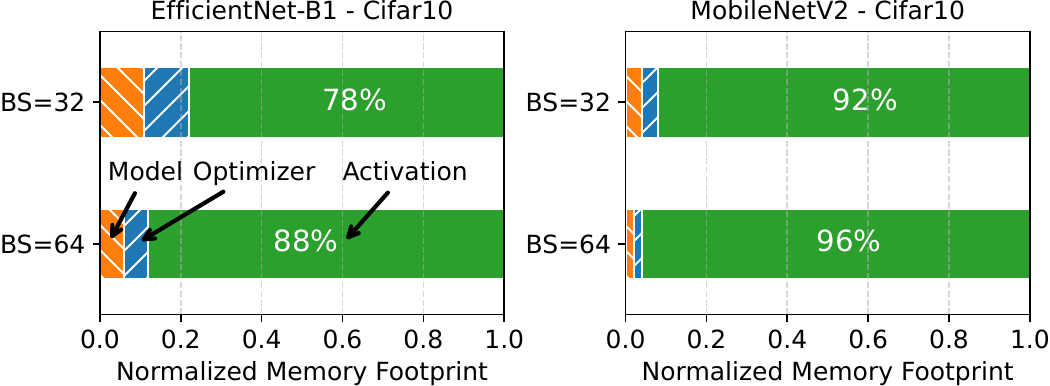}
    \caption{Breakdown of the memory footprint during DNN training, profiled on a Jetson NX.}
    \label{fig:memory-breakdown}
    \vspace{-3pt}
\end{figure}

\subsection{\revise{Hybrid Pipeline Parallelism in Asteroid}}
\label{sec:HPP-workflow}
\revise{\textbf{HPP Architecture and Workflow in Asteroid.} As illustrated in Fig. \ref{fig:stage_partition}, Asteroid's HPP first divides a DNN model into multiple \textbf{stages} where each contains a \textbf{stage model} composed of a set of consecutive network layers. Edge devices in the resource pool will be divided into a corresponding number of \textbf{device groups}, where each contains one or multiple devices. HPP combines pipeline parallelism across these groups with data parallelism within them.
During training, a mini-batch will be split into $M$ smaller \textbf{micro-batches} (with the size of $B$) and injected into the pipeline concurrently to increase parallelism. Micro-batches are broken up further if device groups contain multiple devices.
Each device performs the \textit{forward pass} (\textbf{FP}) and \textit{backward pass} (\textbf{BP}) for the stage model it takes in charge and accumulates gradients for all micro-batches in each mini-batch.
At the end of a mini-batch, gradients in each device group are synchronized using \textbf{AllReduce} and then applied to stage model parameters. The entire mini-batch training process is called an \textbf{HPP-Round}. Fig. \ref{fig:hpp_pipeline} shows a well-designed scheduling arrangement for Asteroid HPP training that models inter-stage network communication, which cannot be neglected due to the significant latency under low-speed links in edge environments. \textbf{Bubbles} are idle periods when a stage waits for data from the previous one, represented by gray blocks.}

\textbf{Memory-efficient 1F1B Micro-batch Scheduling.}
Our memory breakdown experiment in Fig. \ref{fig:memory-breakdown} shows that the peak memory footprint during DNN training can be classified into three categories: (1) model weight memory (including model parameters and accumulated gradients), (2) optimizer memory, and (3) activation memory (intermediate outputs of FP). 
HPP architecture enables each device to only store the corresponding stage model in memory.
In stage $p$, the memory footprints for model weights, optimizer, and activations for a single micro-batch of size $\beta$ are denoted as $\text{Mem}^{(\text{MOD})}_p$, $\text{Mem}^{(\text{OPT})}_p$, and $\text{Mem}^{(\text{ACT})}_p(\beta)$, respectively.

\begin{figure}[t!]
    \setlength{\abovecaptionskip}{0.1cm}
    \setlength{\belowcaptionskip}{-0.4cm}
    \centering
    \includegraphics[width=0.9\linewidth]{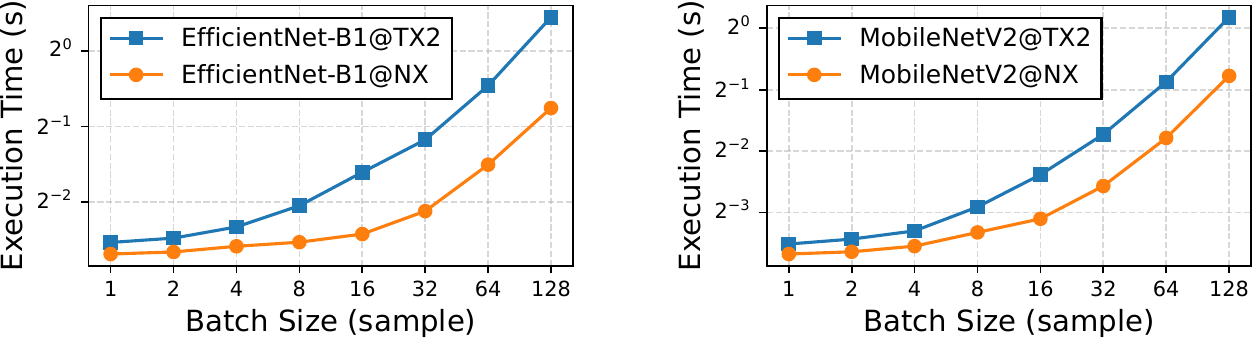}
    \caption{DNN execution time across diverse batch sizes, profiled on Jetson TX2 and Jetson NX.}
    \label{fig:motivation_time}
    \vspace{-5pt}
\end{figure}

The experiment in Fig. \ref{fig:memory-breakdown} shows that the intermediate activations are the main contributor to memory footprint, especially for the CNN-based models which usually have large inter-layer feature maps.
Gpipe \cite{huang2019gpipe} schedules micro-batches in a \textit{backward-after-forward} manner, resulting in a peak memory demand that scales proportionally with the number of concurrently resident micro-batches ($O(M)$), which is memory-unfriendly for edge devices.
Inspired by the idea of gradient accumulation, we adopt a fine-grained micro-batch scheduling that works in a \textit{one-forward-one-backward} (\textbf{1F1B}) manner, which schedules the BP early to release the activation memory produced by FP for reuse.
\revise{We propose performing $K_p$ FP for each stage $p$ before strictly enforcing 1F1B (as illustrated in Figure \ref{fig:hpp_pipeline}, where $K_0=5, K_1=3, K_2=1$), resulting in an activation memory requirement of $O(K_p)$ ($K_p<M$) for each stage $p$. Setting a smaller $K_p$ can reduce memory footprint but compromise stage-level pipeline concurrency. Specifically, with $K_p=1$ for all stages, only one stage will be activated concurrently.
Our experiments in \S \ref{sec:opt-impl} reveal that setting $K_p=2\times(P-p)-1$ ($P$ is the total number of stages) can minimize the peak memory footprint of each stage without sacrificing parallelism efficiency.}
With the above modeling, the total memory footprint of stage $p$ with a micro-batch size $\beta$ in Asteroid is as follows:
\begin{equation}
\vspace{-0.2cm}
\label{equ:mem}
\text{Mem}_p(\beta) = \text{Mem}^{(\text{MOD})}_p + \text{Mem}^{(\text{OPT})}_p + K_p \times \text{Mem}^{(\text{ACT})}_p(\beta).
\vspace{-0.2cm}
\end{equation}

\subsection{Parallelism Planning}
\label{sec:algorithm}
\textbf{DNN Model and Asteroid Profiler.} 
We consider a DNN model as a directed acyclic graph. The graph nodes represents modules like Conv, MaxPool, Attention, etc while the graph edge encodes the data dependency between modules. 
In order to split the model into multiple sequential stages, we topologically sort the graph nodes and transform the DNN model into layers sequence.
Asteroid profiler precisely collects the total output size and weight parameters in bytes for each layer. 
We denote the output activations (and corresponding input gradients) and weight parameters of layer $l$ as $a_l$, and $w_l$, respectively.

We further profile the FP and BP execution time of each layer. 
Existing works \cite{luo2022efficient, jia2022whale} simply assume a linear relationship between batch size and execution time. However, our experiment reveals that smaller batch size fails to fully leverage the parallelism capacity of GPUs, leading to a non-linear correlation between them, as shown in Fig. \ref{fig:motivation_time}. 
Therefore, Asteroid profiler measures the execution time of each layer on realistic edge hardware for various batch sizes. We denote $t_{f}^{d,l}(\beta)$ and $t_{b}^{d,l}(\beta)$ as the FP and BP execution time for layer $l$ on device $d$ using a batch size of $\beta$. We profile the D2D bandwidth between devices $d$ and $d'$ as $b_{d, d'}$.

\begin{table}[t!]
\small
\caption{\revise{Table of Notations for Parallelism Planning}}
\vspace{-0.3cm}
\label{tab:notation}
\begin{tabularx}{\columnwidth}{cX}
\hline
Notation & \multicolumn{1}{c}{Definition} \\
\hline
$L$ & Total number of layers in the DNN model. \\ 
$N$ & Total number of edge devices involved. \\ 
$M$ & Number of micro-batches processed concurrently in a single HPP-Round.  \\ 
$B$ & Size of each micro-batch. \\ 
$P$ & Total number of pipeline stages in the HPP. \\ 
$S$ & Total number of communication and execution steps in the HPP. \\ 
$a_l$ & Size of output activations for the layer $l$. \\ 
$w_l$ & Size of weight parameters for the layer $l$. \\ 
$u_d$ & Memory budget available on device $d$. \\ 
$v_d$ & Computational capacity of device $d$. \\ 
$b_{d,d'}$ & D2D bandwidth between devices $d$ and $d'$. \\ 
$t_{f}^{d,l}(\beta)$ / $t_{b}^{d,l}(\beta)$ & Time to perform FP / BP for layer $l$ on device $d$ with a batch size of $\beta$. \\ 
$T_w^s$ / $T_e^s$ / $T_a^s$ & Execution time for the Waiting / Execution / AllReduce phases in step $s$. \\ 
$E_f^s$ / $E_b^s$ & Time taken for FP / BP in step $s$. \\ 
$\mathcal{G}_s$ & Device group associate with step $s$. \\ 
$\mathcal{D}_s$ & Sub-model associate with step $s$. \\ 
$\mathcal{Y}_s$ & Allocation of micro-batch samples across devices in the group $\mathcal{G}_s$. \\ \hline
\end{tabularx}
\vspace{-0.2cm}
\end{table}

\begin{figure}[t]
    \setlength{\abovecaptionskip}{0.1cm}
    \centering
    \includegraphics[width=0.95\linewidth]{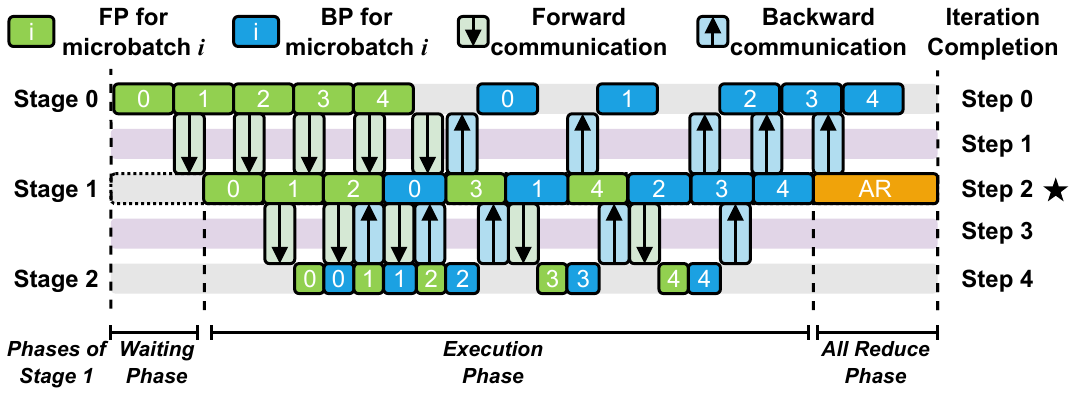}
    \caption{In this example, Step 2 is the dominant step, which is marked with a star.}
    \label{fig:dominate_step}
    \vspace{-15pt}
\end{figure}

\textbf{Optimization Objective Formulation.}
\revise{As illustrated in Fig. \ref{fig:dominate_step}, to emphasize the non-negligible communication latency in edge environments, we abstract the process into separate \textbf{steps}, categorizing them as either \textbf{execution steps} or \textbf{communication steps} for inter-stage communication and stage model execution.}
We divide the training process on each step in an HPP-Round into three phases: \textit{Waiting Phase}, \textit{Execution Phase}, and \textit{AllReduce Phase}, with corresponding times denoted as $T_w^s$, $T_e^s$, and $T_a^s$. 
Our optimization objective is to minimize the \textit{HPP-Round Latency}, which is determined by the step with the largest total latency of three phases:
\begin{equation}
\label{equ:latency}
\begin{aligned}
\text{HPP-Round Latency}=\max_{s \in \{0,1,...,S-1\}}\left(T_w^s+T_e^s+T_a^s\right),
\end{aligned}
\end{equation}
where $S$ denotes the total steps in pipeline. 
For notation simplicity, we denote the forward (backward) execution or communication time of a step $s$ as $E_f^s$ ($E_b^s$), which is associated with a device group $\mathcal{G}_s$ and a sub-model $\mathcal{D}_s$ depending on the pipeline stage of step $s$ ($\mathcal{G}_s$ and $\mathcal{D}_s$ are empty sets for communication steps). The time of \textit{Waiting Phase} and \textit{AllReduce Phase} can be estimated by:
\vspace{-0.2cm}
\begin{equation}
\label{equ:TwTa}
\begin{aligned}
T_w^s=\sum_{i=0}^{s-1} E_f^i, \quad
T_a^s=\frac{2\left(\left|\mathcal{G}_s\right|-1\right) \cdot \sum_{l\in \mathcal{D}_s}w_l}{\left|\mathcal{G}_s\right| \cdot \min _{d, d^{\prime} \in \mathcal{G}_s} b_{d, d^{\prime}}}.
\end{aligned}
\end{equation}
\revise{$T_w^s$ is the sum of FP time till step $s-1$. The size of communication volume per device in $\mathcal{G}_s$ during AllReduce is $\frac{2(|\mathcal{G}_s|-1)}{\mathcal{G}_s} (\sum_{l\in \mathcal{D}_s}w_l)$. The time taken by the AllReduce operation, denoted by $T_a^s$, is further decided by the minimum connection bandwidth among all devices \cite{luo2022efficient, sergeev2018horovod}.}

A remaining hurdle is how to estimate $T_e^s$ of each step. 
As illustrated in Fig. \ref{fig:hybrid_pipeline_parallelism}, the \textit{Execution Phases} of all steps always form a trapezoid shape due to data dependency. Therefore, with an accurate latency estimation of one step, we can infer the $T_e^s$ of the other steps by considering the shift in FP and BP time before and after it. 
We observe that there always exists a step such that the FP and BP are compactly injected in its \textit{Execution Phases}, which is the dominant factor in estimating \textit{Execution Phase} latency.
Thus, we define the step with the fewest number of bubbles during \textit{Execution Phase} as \textbf{dominant step} and its \textit{Execution Phases} latency can be well approximated as the accumulated time of FP and BP of $M$ micro-batches.
Although the dominant step may contain a small fraction of bubbles, leading to an approximation to the true pipeline latency, it has been practically effective in all our evaluations.
As shown in Fig. \ref{fig:dominate_step}, step $2$ is the dominant step without bubble. Conversely, the remaining steps contain scattered bubbles that cannot be accurately estimated, rendering them unsuitable as the dominant step.
Assuming that the index of the dominant step for a pipeline is $dm$, the $T_e^s$ of other steps can be estimated by:
\vspace{-0.1cm}
\begin{equation}
\label{equ:Te}
\begin{aligned}
T_e^s=M \times \left(E_f^{dm}+E_b^{dm}\right)+\left\{
\begin{array}{ll}
\sum_{i=s}^{dm-1}\left(E_f^i+E_b^i\right), &s<dm, \\
-\sum_{i=dm}^{s-1}\left(E_f^i+E_b^i\right), &s \geq dm.
\end{array}\right.
\end{aligned}
\end{equation}

\revise{Based on the aforementioned formulation, once $E_f^s$ and $E_b^s$ are determined, we can calculate the HPP-Round Latency according to Eq. (\ref{equ:latency}). In the following, we first derive and formulate algorithms to estimate $E_f^s$ and $E_b^s$. We then design an innovative dynamic programming algorithm to identify the optimal HPP architecture from all possible configurations, effectively minimizing HPP-Round Latency.}

\begin{algorithm}[t]
\small
\setlength{\textfloatsep}{0.5cm}
\setlength{\floatsep}{0.5cm}
\setlength{\intextsep}{-1em} 
\caption{\revise{Allocation of a micro-batch's samples}}\label{alg:workload-balancing}
\KwIn{$\mathcal{G}_s$: Device Group of step $s$. $B$: Micro-batch Size.}
\KwOut{$\mathcal{Y}_s = \{y_d | d \in \mathcal{G}_s\}$: Micro-batch allocation of step $s$.}
\SetKwRepeat{Do}{do}{while}
\SetKwFunction{MAB}{MemoryAwareBalancing}
\SetKwFunction{SWO}{StragglerWorkloadOffloading}
\SetKw{Return}{Return}
\SetKwProg{Ft}{Function}{:}{}
\Ft{\MAB{$\mathcal{G}, \beta$}}{
    \uIf{$|\mathcal{G}| == 0$ and $\beta > 0$}{\textbf{Exit with Fail}; $\quad \triangleright$ Set $T(i\rightarrow j, \mathcal{G})$ as $\infty$}
    \ElseIf{$\beta == 0$}{\Return\;}
    \ForEach{$d \in \mathcal{G}$}{
        ${bs}_d \leftarrow$ Maximum batch size of $d$ under budget $u_d$\;
        $y_d \leftarrow y_d + \min(\frac{v_d}{\sum_{d\in \mathcal{G}}v_d} \beta, bs_d)$\;
        Update $u_d$ with remaining memory budget\;
    }
    $\mathcal{G}' \leftarrow$ Devices in $\mathcal{G}$ with remaining memory\;
    $\beta' \leftarrow$ Unallocated batch size in $\beta$\;
    \MAB{$\mathcal{G}', \beta'$}\;
}
\Ft{\SWO{}}{
Sort devices within $\mathcal{G}_s$ in ascending order based on their execution latency with $\mathcal{Y}_s$\;
\Do{$new\_straggler$ faster than $old\_straggler$}{
$old\_straggler \leftarrow$ The slowest device in $\mathcal{G}_s$\;
  Transfer a block of samples from the straggler to the fastest device with sufficient memory\;
  Reorder devices within $\mathcal{G}_s$ in ascending\;
  $new\_straggler \leftarrow$ The slowest device in $\mathcal{G}_s$\;
}
}
\MAB{$\mathcal{G}_s, B$}; $\quad \triangleright$ Phase 1\\
\SWO{}; $\quad \triangleright$ Phase 2
\end{algorithm}

\textbf{\revise{Estimating $E_f^s$ and $E_b^s$ for step $s$.}}
For the communication steps, $E_f^s$ and $E_b^s$ can be estimated by the size of the intermediate tensors required for transmission between stages and the profiled D2D communication bandwidth. 
\revise{In the following, we concentrate on estimating $E_f^s$ and $E_b^s$ for execution steps, with a key objective being the optimal allocation of a micro-batch's samples among resource-diverse devices. This allocation seeks to minimize data parallel execution time within the memory budget of each device.
We denote $T(i\rightarrow j, \mathcal{G}_s)$ as the optimal time taken by device group $\mathcal{G}_s$ spanning layers $i$ through $j$ for both FP and BP. The optimization target can be formulated as follows:}
\vspace{-0.3cm}
\begin{equation}
\label{equ:constrain}
\begin{aligned}
T(i\rightarrow j, \mathcal{G}_s) = \min_{y_d \in \mathcal{Y}_s}\max_{d \in \mathcal{G}_s}\sum_{l=i}^j[t_{f}^{d,l}(y_d) + t_{b}^{d,l}(y_d)],\\
s.t. \sum_{d\in \mathcal{G}_s} y_d = \text{B}, \quad \text{Mem}_s(y_d) \leq u_d,\\
\end{aligned}
\end{equation}
\revise{where $\mathcal{Y}_s = \{y_d | d \in \mathcal{G}_s\}$ defines the set of the allocation of a micro-batch across devices in step $s$, with each element $y_d$ representing the number of samples allocated to device $d$. 
The peak memory footprint is given by $\text{Mem}_s(y_d)$ (refer to Eq. (\ref{equ:mem})). $u_d$ denotes the memory budget of device $d$.
$E_f^s$ and $E_b^s$ can be estimated as follows once the $\mathcal{Y}_s$ is determined.}
\vspace{-0.3cm}
\begin{equation}
    \label{equ:Efs-Ebs}
    E_f^s = \max_{d\in \mathcal{G}_s} \sum_{l=i}^j t_f^{d,l}(y_d),\quad E_b^s = \max_{d\in \mathcal{G}_s} \sum_{l=i}^j t_b^{d,l}(y_d).
\end{equation}
\vspace{-0.2cm}

To efficiently solve the aforementioned constrained optimization problem, we propose an iterative planning algorithm as outlined in Algorithm \ref{alg:workload-balancing}. The algorithm can be divided into two distinct phases: \textit{memory-aware workload balancing phase} and  \textit{workload offloading phase}.
In the first phase, we recursively distribute the workload among devices in a manner that balances the workload based on their computing capacities, while strictly adhering to the memory budgets (line 1-12). A device's computing capacity $v_d$ is defined as the inverse of the sum of FP and BP execution latency with a micro-batch:
\vspace{-0.2cm}
\begin{equation}
\label{equ:capacity}
v_d=\left(\sum_{l=i}^{j}[t_{f}^{d,l}(B) + t_{b}^{d,l}(B)]\right)^{-1}.
\end{equation}

\begin{figure}[t]
    \setlength{\abovecaptionskip}{0.1cm}
    \setlength{\belowcaptionskip}{-0.5cm}
    \centering
    \includegraphics[width=0.95\linewidth]{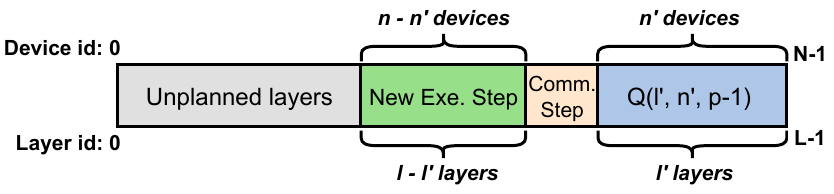}
    \caption{An instance of dynamic programming planning process for $\widetilde{Q}(l,n,l',n',p-1)$.}
    \label{fig:dynamic-programming}
    \vspace{-0.1cm}
\end{figure}

As previously stated, our experiment reveals the non-linear relationship between batch size and execution time. Consequently, relying solely on the first phase will lead to suboptimal outcomes. 
In the second phase, we iteratively transfer the sample workload (one block at a time) from the straggler (the slowest device) to the fastest device with sufficient memory. The iteration terminates when the workload offloading results in a slower straggler (line 13-20).
We can adjust the \textit{block size} according to the micro-batch size to trade-off between planning overhead and balancing. 

\textbf{\revise{Dynamic Programming HPP Planning.}}
Our dynamic programming algorithm searches for optimal HPP configurations to minimize HPP-Round Latency.
To reduce the searching complexity for orchestration across heterogeneous devices, we sort the devices in descending order by their memory capacity and map stages accordingly. This sorting is inspired by the observation in \S \ref{sec:opt-impl} that
the earlier stages in Asteroid typically require more storage space for activations compared to the later stages.

We consider a DNN model consisting of $L$ layers and aim to leverage the computing resource from an edge resource pool with $N$ edge devices to efficiently perform HPP training. To achieve this, a novel \textit{dynamic programming algorithm} is devised which facilitates optimal parallelism planning.
We denote $Q(l,n,p)$ as the \textit{HPP-Round Latency} of the optimal pipeline planning when slicing the last $l$ consecutive layers into $p$ stages and putting them onto the last $n$ devices. 
The goal of our algorithm is to calculate $\min_p Q(L,N,p)$. $Q(l,n,p)$ is updated as follows:
\begin{equation}
\label{equ:Q}
\begin{aligned}
Q(l,n,p)=\min_{l',n'}\widetilde{Q}(l,n,l',n',p-1), 
\end{aligned}
\end{equation}
where $\widetilde{Q}(l,n,l',n',p-1)$ represents the \textit{HPP-Round Latency} of a pipeline which comprises two parts (see Fig. \ref{fig:dynamic-programming} as illustration): (1) an optimal sub-pipeline $Q(l',n',p-1)$ consisting of the last $l'$ layers with $p-1$ stages cross the last $n'$ devices. (2) a new single stage (execution step) with layers $L-l$ to $L-l'$ replicated over remaining $n-n'$ devices using DP.

To obtain $\widetilde{Q}(l,n,l',n',p-1)$, its dominant step must first be determined.
Finding the step with the fewest bubbles during \textit{Execution Phase} is equivalent to finding the step with the largest total FP and BP execution time after alignment. 
Consequently, we align and compare the total execution or communication time of the dominant step of sub-pipeline $Q(l',n',p-1)$, the new execution step at the head, and the communication step between the first step and $Q(l',n',p-1)$.
The largest one contains the fewest bubbles during \textit{Execution Phase} will be updated as the new dominant step:
\begin{equation}
\label{equ:dm}
\max
\begin{cases}
M \times\left(E_f^{d m^*}+E_b^{dm^*}\right)+\sum_{i=0}^{d m^*+1}\left(E_f^i+E_b^i\right), \\
M \times\left(E_f^{ne}+E_b^{ne}\right), \\
M \times \left(E_f^{nc}+E_b^{nc}\right)+(E_f^{ne}+E_b^{ne}),
\end{cases}
\end{equation}
where $dm^*$ denotes the original index of dominant step in sub-pipeline $Q(l',n',p-1)$, and $ne$ and $nc$ represent the index of the new execution step and new communication step, respectively. After determining the dominant step, we can estimate $T_w$, $T_e$ and $T_a$ for each step according to Eq. (\ref{equ:TwTa}) and \ref{equ:Te}, and then infer the \textit{HPP-Round Latency} by Eq. (\ref{equ:latency}).

\begin{algorithm}[t]
\small
\setlength{\textfloatsep}{0.5cm}
\setlength{\floatsep}{0.5cm}
\setlength{\intextsep}{-1em} 
\caption{\revise{Dynamic Programming HPP Planning}}\label{alg:dynamic_programming}
\SetKwRepeat{Do}{do}{while}
\SetKwFunction{MAB}{MemoryAwareBalancing}
\SetKwFunction{SWO}{StragglerWorkloadOffloading}
\SetKw{Return}{Return}
\SetKwProg{Ft}{Function}{:}{}
\For{p \textbf{from} 1 \textbf{to} $min(L,N)$}{
    \For{n \textbf{from} 1 \textbf{to} N}{
        \For{l \textbf{from} 1 \textbf{to} L}{
            \For{n' \textbf{from} 0 \textbf{to} n}{
                \For{l' \textbf{from} 0 \textbf{to} l}{
                    Get $E_f^s$ and $E_b^s$ with Alg. \ref{alg:workload-balancing} and Eq. (\ref{equ:Efs-Ebs})\;
                    Update Dominant Step with Eq. (\ref{equ:dm})\;
                    Get $T_w^s$, $T_e^s$ and $T_a^s$ with Eq. (\ref{equ:TwTa}) and (\ref{equ:Te})\;
                    Get HPP-Round Latency with Eq. (\ref{equ:latency})\;
                }
            }
            Update $Q(l,n,p)$ with Eq. (\ref{equ:Q})\;
        }
}
}

\end{algorithm}

\revise{We summarize our dynamic programming planning in Algorithm \ref{alg:dynamic_programming}. During the execution of algorithm, we concurrently record the parallel configurations of $Q(l,n,p)$, which includes model partitioning points, device grouping strategies, and micro-batch allocations within groups. This process identifies the optimal parallel configurations for HPP upon completion.}
In \S \ref{sec:complexity}, we demonstrate the efficiency and lightweight nature of our HPP planning, illustrating its suitability for deployment in edge environments.

\subsection{Fault-Tolerant Pipeline Replay}
\label{sec:replay}
Devices at the edge exhibit strong dynamics, as they may leave training at any time, or disconnect due to energy depletion or network anomalies. Such single-device failures can cause the following issues: (1) The device departing can result in the loss of the trained weights. (2) An abnormal device in pipeline can lead to blockages, which necessitate pipeline re-planning. A straw-man proposal is to aggregate stage models to coordinator, rerun the planning algorithm, and redistribute weights based on the new configuration. However, this \textit{heavy rescheduling} may induce considerable latency in the re-planning and model transmission.
To tackle these issues, Asteroid incorporates an on-the-fly fault-tolerant pipeline-replay \textit{lightweight rescheduling}, featuring three modules that efficiently respond to resource fluctuations:

\begin{figure}[t]
    \setlength{\abovecaptionskip}{0.1cm}
    \setlength{\belowcaptionskip}{-0.5cm}
    \centering
    \includegraphics[width=0.9\linewidth]{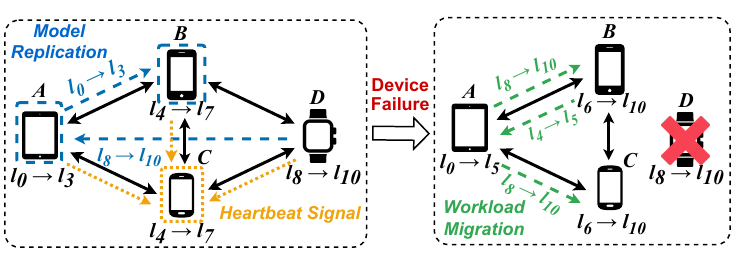}
    \caption{An instance of fault-tolerant pipeline replay.}
    \label{fig:pipeline_replay_example}
\end{figure}

\noindent\textbf{1. Heartbeat-guided Failure Detection.} As illustrated in Fig. \ref{fig:pipeline_replay_example}(Left), each device periodically emits heartbeat signals to the central coordinator as proof of liveness. The absence of these signals within an expected timeframe may indicate a potential device failure. The coordinator will further dispatch a probe message specifically to the suspected device for confirmation of its operational status.

\noindent\textbf{2. Topology-driven Model Replication.} Asteroid adopts a topology-driven model replication to mitigate the loss of trained weights due to device failure. As depicted in Fig. \ref{fig:pipeline_replay_example}(Left), single-device stage (A and D) periodically backs up its stage model to a dedicated device (\textit{backup node}) in the next stage, with the last stage being backed up to the first stage. The model weights can be restored from backup node when device failure occurs. 
In the presence of device failure in multi-device stages (B and C), model weights can be restored from other surviving devices within the same stage.
We note that our topology-driven replication can complement other fault-tolerance mechanisms to address more dynamic scenarios, such as the simultaneous failure of multiple devices.

\noindent\textbf{3. Layer-wise Lightweight Pipeline Re-planning.}
To expedite the pipeline re-planning, we employ a layer-wise lightweight approach as a substitute for rerunning the entire planning algorithm. 
We quantify the workload by profiling the FLOPs of all model layers. In the case of a device failure, the associated workload is reallocated among the remaining stages based on their computing capacity ($\sum_{d \in G_s} v_d$). This can be achieved by making a minor adjustment to the layer partitioning points of the current planning configuration. As illustrated in Fig. \ref{fig:pipeline_replay_example}(Right), our mechanism facilitates concurrent layer migration between adjacent stages according to the updated configuration, maximizing bandwidth utilization while minimizing redundant weights transmission.

\textbf{Putting It All Together.} An illustrative example with four devices is presented in Fig. \ref{fig:pipeline_replay_example}. Devices \textit{A}, \textit{B}, and \textit{D} emit periodic heartbeat signals to coordinator \textit{C} to confirm their liveness. Device \textit{A} periodically checkpoints its stage model to backup node \textit{B}, while Device \textit{D} checkpoints its stage model to \textit{A}. When detecting a failure in Device \textit{D}, our lightweight FLOPs-based approach recalibrates the original layer partitioning. Subsequently, all stages concurrently migrate layers based on the new partitioning points. The layer weights initially on failed device \textit{D} can be restored from backup node \textit{A}. After re-planning, a refined two-stage pipeline involving three devices will take over the collaborative model training.

\begin{figure}[t]
    \setlength{\abovecaptionskip}{0.1cm}
    \setlength{\belowcaptionskip}{-0.5cm}
    \centering
    \begin{minipage}[b]{0.375\linewidth}
        \includegraphics[width=\linewidth]{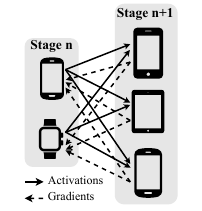}
        \caption{A case of stage replication.}
        \label{fig:repli}
    \end{minipage}
    \hfill
    \begin{minipage}[b]{0.565\linewidth}
        \includegraphics[width=\linewidth]{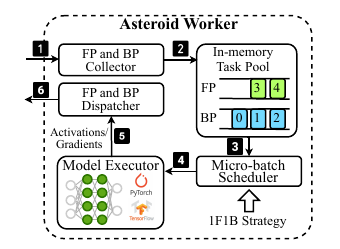}
        \caption{An instance of Asteroid Worker.}
        \label{fig:worker}
    \end{minipage}
\end{figure}

\begin{table*}[t]
\small
\setlength{\textfloatsep}{0.5cm}
\setlength{\floatsep}{0.5cm}
\setlength{\intextsep}{-1em} 
    \caption{Summary of throughput results comparing Asteroid with on-device training, data parallelism (DP), and pipeline parallelism (PP). The pipeline configuration generated by Asteroid is visualized in Fig. \ref{fig:configuration}. We select the most powerful device in each edge environment as the platform for on-device training.}
    \vspace{-10pt}
    \label{table:throughput}
    \centering
    \begin{tabular}{ccccrccrc}
    \hline
    \multirow{2}{*}{Task}                                                            & \multirow{2}{*}{Model}           & \multirow{2}{*}{Dataset}        & \multirow{2}{*}{Input Size}                  & \multicolumn{1}{c}{\multirow{2}{*}{\begin{tabular}[c]{@{}c@{}}Edge\\ Environment\end{tabular}}} & \multirow{2}{*}{\begin{tabular}[c]{@{}c@{}}Asteroid \\ Config.\end{tabular}} & \multicolumn{3}{c}{\begin{tabular}[c]{@{}c@{}}Speedup over\end{tabular}} \\ \cline{7-9} 
                                                                                     &                                  &                                 &                                              & \multicolumn{1}{c}{}                                                                           &                                                                                  & Device           & \multicolumn{1}{c}{DP}        & \multicolumn{1}{c}{PP}        \\ \hline
    \multirow{9}{*}{\begin{tabular}[c]{@{}c@{}}Image \\ Classi-\\fication\end{tabular}} & \multirow{3}{*}{EfficientNet-B1 \cite{tan2019efficientnet}} & \multirow{3}{*}{Cifar-10 \cite{cifar10}}       & \multirow{3}{*}{3 $\times$ 32 $\times$ 32}   & A (100Mbps)                                                                                      & \ding{182}                                                                                & 4.4$\times$         & 2.1$\times$                   & 2.8$\times$                      \\
                                                                                     &                                  &                                 &                                              & B (100Mbps)                                                                                      & \ding{185}                                                                                 & 3.0$\times$         & 4.8$\times$                   & 9.7$\times$                      \\
                                                                                     &                                  &                                 &                                              & B (1000Mbps)                                                                                  & \ding{185}                                                                                 & 3.7$\times$         & 2.1$\times$                   & 1.4$\times$                      \\ \cline{2-9} 
                                                                                     & \multirow{3}{*}{MobileNetV2 \cite{sandler2018mobilenetv2}}     & \multirow{3}{*}{Cifar-10 \cite{cifar10}}       & \multirow{3}{*}{3 $\times$ 32 $\times$ 32}   & A (100Mbps)                                                                                       & \ding{183}                                   & 4.5$\times$         & 1.5$\times$                   & 3.5$\times$                      \\
                                                                                     &                                  &                                 &                                              & B (100Mbps)                                                                                     & \ding{186}                                                                                 & 3.2$\times$         & 2.3$\times$                   & 11.2$\times$                      \\
                                                                                     &                                  &                                 &                                              & B (1000Mbps)                                                                                 & \ding{187}                                                                                 & 3.8$\times$         & 1.2$\times$                   & 1.3$\times$                      \\ \cline{2-9} 
                                                                                     & \multirow{3}{*}{ResNet50 \cite{he2016deep}}        & \multirow{3}{*}{Mini-ImageNet \cite{vinyals2016matching}}       & \multirow{3}{*}{3 $\times$ 224 $\times$ 224} & A (100Mbps)                                                                                       & \ding{183}                           & 3.4$\times$         & 3.6$\times$                   & 5.8$\times$                      \\
                                                                                     &                                  &                                 &                                              & B (100Mbps)                                                                                     & \ding{187}                                                                                 & 1.5$\times$         & 6.1$\times$                   & 12.2$\times$                      \\
                                                                                     &                                  &                                 &                                              & B (1000Mbps)                                                                                 & \ding{185}                                                                                 & 3.7$\times$         & 2.9$\times$                   & 3.1$\times$                      \\ \hline
    \multirow{3}{*}{\begin{tabular}[c]{@{}c@{}}Language \\ Model\end{tabular}}                                                  & \multirow{3}{*}{Bert-small \cite{devlin2018bert}}      & \multirow{3}{*}{Synthetic Data} & \multirow{3}{*}{32 $\times$ 512}             & A (100Mbps)                                                                                       & \ding{184}           & 3.5$\times$         & 6.4$\times$                   & 1$\times$                        \\
                                                                                     &                                  &                                 &                                              & B (100Mbps)                                                                                     & \ding{188}                                                                               & 1.3$\times$         & 6.8$\times$                   & 1$\times$                        \\
                                                                                     &                                  &                                 &                                              & B (1000Mbps)                                                                                 & \ding{188}                                                                                & 3.9$\times$         & 4.2$\times$                   & 1.3$\times$                      \\ \hline
    \end{tabular}
    \vspace{-10pt}
    \end{table*}

\begin{figure*}[t]
    \setlength{\abovecaptionskip}{0.1cm}
    \setlength{\belowcaptionskip}{-0.3cm}
    \centering
    \subfigure[Configuration \ding{182}.]{
        \begin{minipage}[t]{0.12\linewidth}
        \centering
        \includegraphics[height=1.7cm]{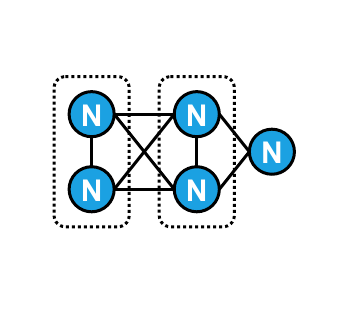}
        \label{fig:config1}
        \end{minipage}
    }
    \subfigure[Configuration \ding{183}.]{
        \begin{minipage}[t]{0.14\linewidth}
        \centering
        \includegraphics[height=1.7cm]{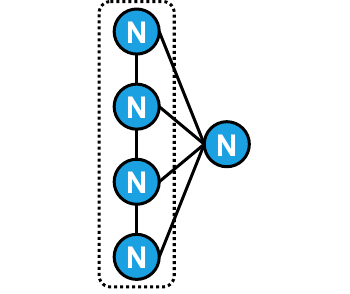}
        \label{fig:config2}
        \end{minipage}
    }
    \subfigure[Configuration \ding{184}.]{
        \begin{minipage}[t]{0.12\linewidth}
        \centering
        \includegraphics[height=1.7cm]{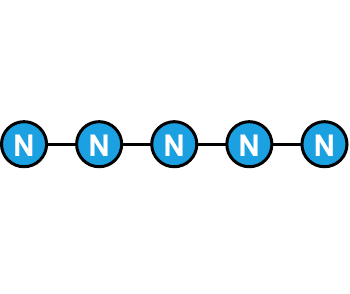}
        \label{fig:config3}
        \end{minipage}
    }
    \subfigure[Configuration \ding{185}.]{
        \begin{minipage}[t]{0.12\linewidth}
        \centering
        \includegraphics[height=1.7cm]{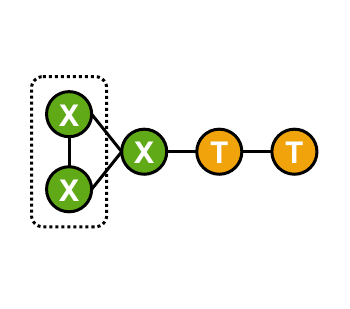}
        \label{fig:config4}
        \end{minipage}
    }
    \subfigure[Configuration \ding{186}.]{
        \begin{minipage}[t]{0.12\linewidth}
        \centering
        \includegraphics[height=1.7cm]{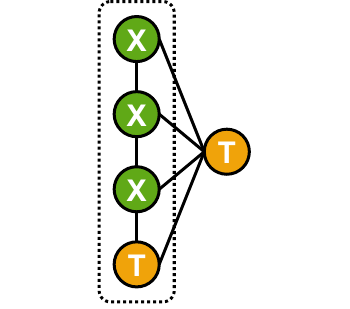}
        \label{fig:config5}
        \end{minipage}
    }
    \subfigure[Configuration \ding{187}.]{
        \begin{minipage}[t]{0.12\linewidth}
        \centering
        \includegraphics[height=1.7cm]{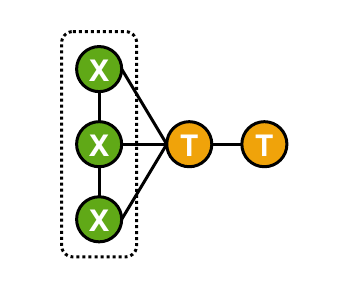}
        \label{fig:config6}
        \end{minipage}
    }
    \subfigure[Configuration \ding{188}.]{
        \begin{minipage}[t]{0.12\linewidth}
        \centering
        \includegraphics[height=1.7cm]{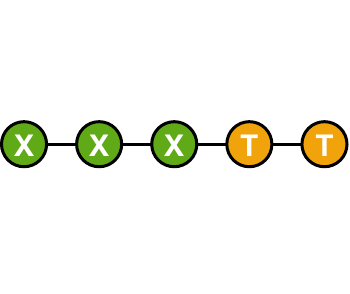}
        \label{fig:config7}
        \end{minipage}
    }
    \vspace{-0.1cm}
    \caption{HPP configurations for the experiments in Table \ref{table:throughput}. The dashed box indicates a pipeline stage where the devices inside perform data parallelism. "N", "T", and "X" indicate Jetson Nano, TX2, and NX, respectively.}
    \label{fig:configuration}
\end{figure*}

\section{Implementation}
We have fully implemented a prototype system of Asteroid with $\sim$2,000 LOC in Go and Python in total atop PyTorch \cite{pytorch}. Although we use PyTorch for auto-differentiation and computation graph execution, Asteroid is extensible and can work well with other lightweight ML frameworks such as TF-Lite \cite{tflite} and MNN \cite{jiang2020mnn}.

\begin{table}[t!]
\small
\setlength{\abovecaptionskip}{0cm}
    \setlength{\belowcaptionskip}{-0.1cm}
\caption{Specifications of edge devices in experiments.}
\begin{tabular}{ccc}
\hline
\textbf{Edge Device} & \textbf{GPU Processor}     & \textbf{\begin{tabular}[c]{@{}c@{}}Memory\end{tabular}} \\ \hline
Jetson Nano \cite{jetson-nano}         & 128-core NVIDIA Maxwell   & 4GB                                                               \\ 
Jetson TX2 \cite{jetson-TX2}         & 256-core NVIDIA Pascal & 8GB                                                               \\ 
Jetson NX \cite{jetson-NX}           & 384-core NVIDIA Volta  & 8GB                                                               \\ \hline
\end{tabular}
\label{tab:hardware}
\vspace{-10pt}
\end{table}

\begin{table}[t!]
\small
\caption{Heterogeneous edge env. used in experiments.}
\vspace{-0.4cm}
\begin{tabular}{cc|cc}
\hline
\textbf{ID} & \textbf{Devices}                                                        & \textbf{ID} & \textbf{Devices}                                                                           \\ \hline
A            & 5 $\times$ Nano                                                         & C            & \begin{tabular}[c]{@{}c@{}}1 $\times$ NX, 2 $\times$ TX2, 3 $\times$ Nano\end{tabular} \\ 
B            & \begin{tabular}[c]{@{}c@{}}3 $\times$ NX, 2 $\times$ TX2\end{tabular} & D            & \begin{tabular}[c]{@{}c@{}}1 $\times$ TX2, 3 $\times$ Nano\end{tabular}                  \\ \hline
\end{tabular}
\label{tab:iot-cluster}
\vspace{-10pt}
\end{table}

\textbf{Stage Replication.}
Fig. \ref{fig:repli} illustrates an example where stage $n$ is replicated onto two devices and stage $n+1$ spans three devices. During FP, $1/3$ of the activations derived by each device on stage $n$ is sent to each device on stage $n+1$. During BP, each device on stage $n+1$ splits its gradients into two sets and sends back to each device on stage $n$. We ensure that FP and BP of each sample are co-located on one device. We use PyTorch’s \textit{Distributed-DataParallel} library \cite{pytorch-ddp} to synchronize parameters for data-parallel stages.

\textbf{Micro-batch FP and BP Scheduling.} 
Each device deploys an \textit{Asteroid Worker} to manage micro-batch FP and BP scheduling, as depicted in Fig. \ref{fig:worker}. Workers asynchronously receive activations or gradients from the preceding stage  (\squared{1}), packaging them into FP and BP tasks and collecting in an in-memory task pool (\squared{2}). 
Micro-batch scheduler fetches tasks for FP/BP execution (\squared{3}), adhering to a predefined scheduling strategy (e.g., 1F1B) (\squared{4}).
Intermediate tensors derived from FP/BP execution are passed to the dispatcher (\squared{5}) for asynchronous transmission to the next stage (\squared{6}).

\vspace{-0.3cm}

\section{Evaluation}
\label{sec:eval}

\subsection{Experimental Setup}
\textbf{Models and Datasets.} We evaluate Asteroid with 4 typical DNN models that are widely used in computer vision (i.e., EfficientNet-B1 \cite{tan2019efficientnet}, MobileNetV2 \cite{sandler2018mobilenetv2}, ResNet-50 \cite{he2016deep}) and natural language processing (i.e., Bert-small \cite{devlin2018bert}). We use Cifar-10 \cite{cifar10} dataset with input size $3 \times 32 \times 32$ for both EfficientNet-B1 and MobileNetV2, and Mini-ImageNet  \cite{vinyals2016matching} dataset with input resized to $3 \times 224 \times 224$. For Bert-small, we build a synthetic dataset with input shape $32 \times 512$.

\textbf{Heterogeneous Edge Environment Setup.} We use 3 heterogeneous off-the-shelf edge devices listed in Table \ref{tab:hardware} in our experiments.  
We analyze our system performance on four edge environments (Table \ref{tab:iot-cluster}) where each is a different combination of heterogeneous edge devices. We consider both 100Mbps and 1000Mbps intra-cluster network bandwidth setting to simulate different network conditions in real edge environments.

\begin{figure*}[t]
    \setlength{\abovecaptionskip}{0.1cm}
    \setlength{\belowcaptionskip}{-0.2cm}
    \centering
    \subfigure[Training throughput compared with existing approaches on Env. B.]{
        \begin{minipage}[t]{0.485\linewidth}
        \centering
        \includegraphics[width=\linewidth]{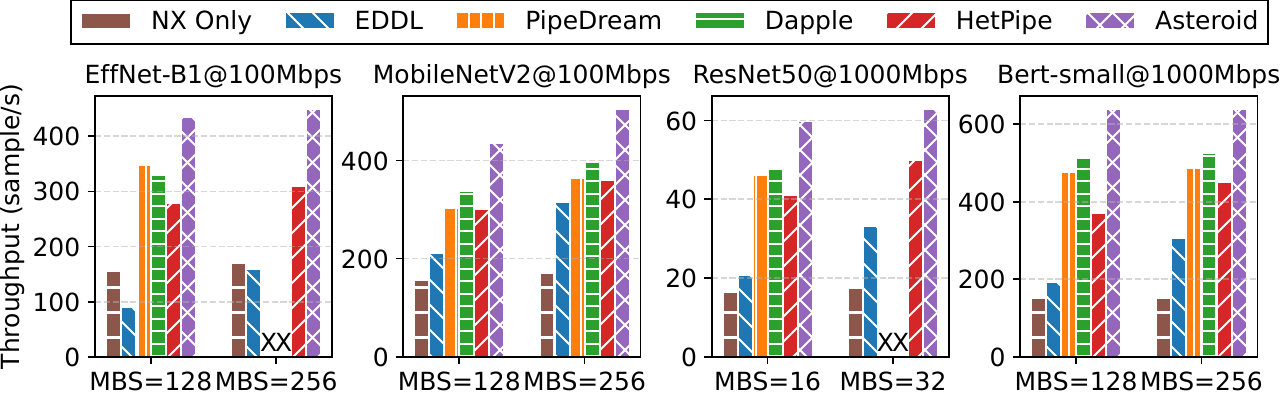}
        \label{fig:MBS2throughput_1}
        \vspace{-10pt}
        \end{minipage}
    }
    \subfigure[Training throughput compared with existing approaches on Env. C.]{
        \begin{minipage}[t]{0.485\linewidth}
        \centering
        \includegraphics[width=\linewidth]{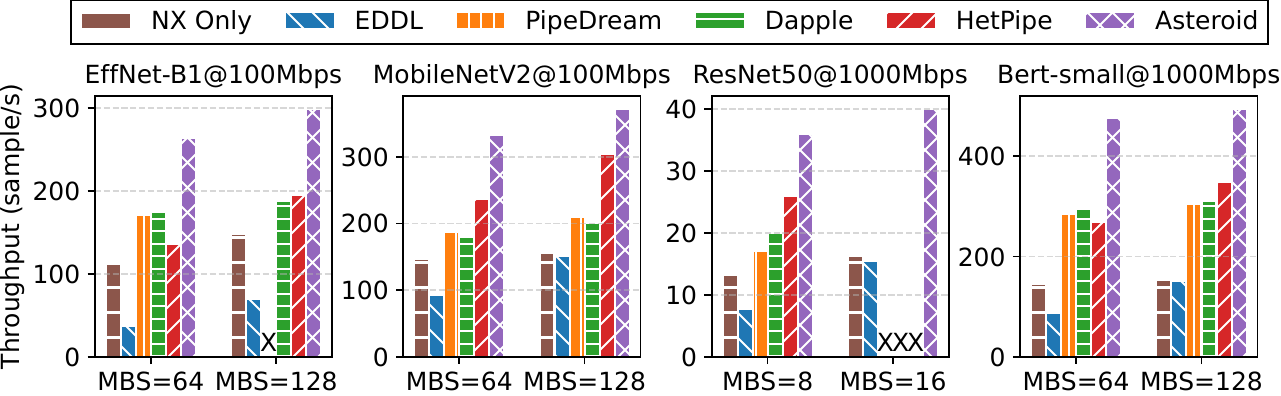}
        \label{fig:MBS2throughput_2}
        \vspace{-10pt}
        \end{minipage}
    }
    \caption{Training throughput comparison under various settings. $\times$ means out-of-memory error.}
    \label{fig:MBS2throughput}
\end{figure*}

\begin{figure*}[t]
    \centering
    \setlength{\abovecaptionskip}{0.1cm}
    \includegraphics[width=0.99\linewidth]{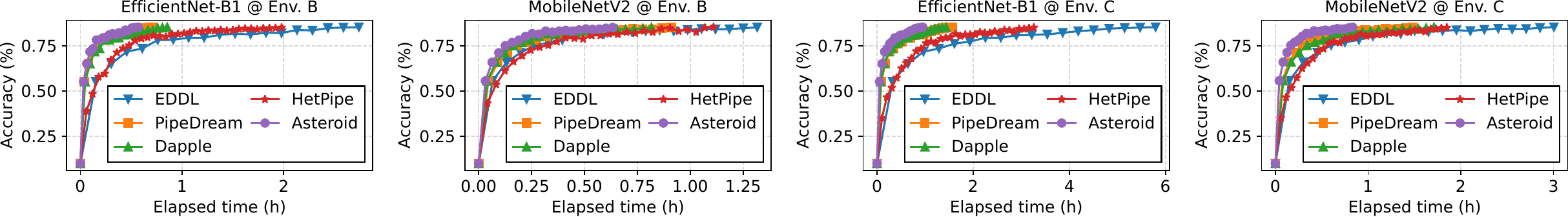}
    \caption{Training convergence of EfficientNet-B1 and MobileNetV2 on Env. B and C compared with baselines.}
    \label{fig:time2acc}
    \vspace{-10pt}
\end{figure*}

\textbf{Baseline Methods.} We implement and compare Asteroid with both the widely-used traditional baselines and the state-of-the-art parallel training methods:
\vspace{-0.2cm}
\begin{itemize}[leftmargin=*]
    \item \textbf{Data Parallelism (DP)} \cite{li2014communication} is a conventional parallel training method that distributes batch data across cluster devices for concurrent processing.
    \item \textbf{Pipeline Parallelism (PP)} \cite{huang2019gpipe} is a conventional parallel training approach that divides the DNN model into sequential stages and processes them in a pipeline manner.
    \item \textbf{EDDL} \cite{hao2021eddl} is a collaborative edge training system that leverages data parallelism on edge device clusters.
    \item \textbf{PipeDream} \cite{narayanan2019pipedream} explores the HPP for asynchronous training in homogeneous accelerator clusters within datacenters. We compare our planning algorithm with PipeDream's under synchronous training scenarios.
    \item \textbf{Dapple} \cite{fan2021dapple} devises an HPP method dedicated to the synchronous DNN training in large-scale homogeneous accelerators cluster in datacenters.
    \item \textbf{HetPipe} \cite{park2020hetpipe} facilitates asynchronous parallel training of large DNN models on heterogeneous GPU clusters by treating sub-groups of GPUs as virtual workers and employing intra-worker PP and inter-worker DP.
\end{itemize}

\vspace{-0.5cm}
\subsection{Comparison with DP and PP}
Table \ref{table:throughput} summarizes the training throughput results comparing Asteroid with on-device, DP and PP training methods. 
To facilitate a better comparison, we implement heterogeneous workload balancing for both DP and PP, and further employ our 1F1B scheduling for PP.
\revise{To evaluate Asteroid's performance across clusters with diverse computational capabilities and network bandwidths, we conduct experiments on three different edge environments for each model: Env. A, Env. B and Env. B with 1000Mbps D2D bandwidth.} For a fair comparison, all methods for each model share a same global mini-batch size of $2048$ for EfficientNet-B1, MobileNetV2, Bert-small and $256$ for ResNet50.
While prior works have focused on optimizing DP and PP, our evaluation results indicate that Asteroid's HPP is the best for complex edge environment.
In CNN-based models, feature map size decreases as layers deepen, and most parameters are in the end's fully connected layers. To reduce inter-stage communication with large activations and minimize AllReduce overhead, Asteroid employs DP in earlier layers and PP in later layers. For Transformer-based language models with huge parameters and small inter-layer activations, Asteroid's planner suggests a straight pipeline.
As shown in Table \ref{table:throughput}, Asteroid achieves a training speedup of $2.1\times$-$6.8\times$ and $1.3\times$-$12.8\times$ compared to DP and PP methods, respectively. Specifically, for ResNet50 on Env. B, Asteroid achieves up to $12.2 \times$ faster than PP. We investigate the underlying reasons and find that the inter-stage communication latency between the first and second stage is $24 \times$ greater than the execution time of the first stage, which dominates the entire training process. In particular, Asteroid achieves nearly linear scaling and is $3.9\times$ faster than single Jetson NX on the BERT-small model.

\subsection{Comparison with Existing Systems}
\textbf{Training Throughput Performance.}
We compare Asteroid with other state-of-the-art distributed training approaches. \revise{To showcase Asteroid’s robustness in heterogeneous environments, we conducted experiments for all four models in two edge environments, B and C, each exhibiting different levels of heterogeneity.}
The results are demonstrated in Fig. \ref{fig:MBS2throughput}.
Our key observation is that Asteroid consistently and remarkably outperforms other existing parallelism under heterogeneous and resource-constrained edge environments.
Specifically, Asteroid surpasses the DP method EDDL, achieving a throughput increase of $1.6\times$-$6.9\times$. When compared to HPP methods PipeDream and Dapple, Asteroid attains throughput improvements of $1.3\times$-$2.1\times$ and $1.2\times$-$1.8\times$, respectively. This is because both PipeDream and Dapple are designed for homogeneous accelerator clusters in datacenter, which results in unbalanced and suboptimal workload partition and device grouping for both inter-stage and intra-stage. 
\revise{When compare to HetPipe, Asteroid achieves a throughput increase of $1.2\times$-$1.9\times$. HetPipe considers device heterogeneity but requires full model gradient exchange after each training iteration, leading to increased D2D communication volume. Also, HetPipe necessitates a centralized parameter server (PS) for asynchronous gradient exchange. Utilizing a bandwidth-limited edge device as a PS can become a bottleneck in the distributed system, and may also disrupt the communication essential for HPP training in which it participates.}
This experiment further reveals that our planning algorithm effectively balances training throughput and peak memory footprint. Conversely, PipeDream, Dapple, and HetPipe do not consider device memory budget in their parallelism planning algorithms, leading to OOM errors with their planning configurations.

\begin{figure}[t!]
    \setlength{\abovecaptionskip}{0.1cm}
    \setlength{\belowcaptionskip}{-0.5cm}
    \centering
    \subfigure[Ablation study of planning algorithm. A: inter-stage planning. B: intra-stage planning.]{
        \includegraphics[width=0.38\linewidth]{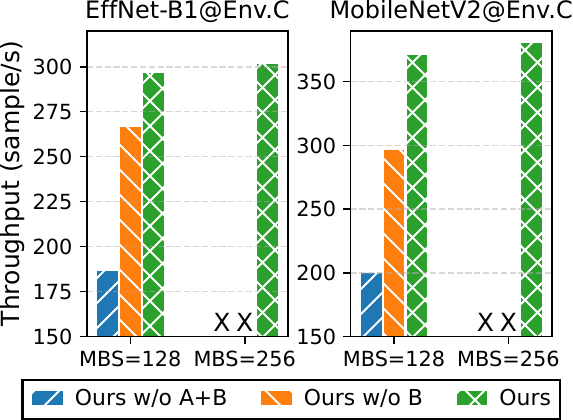}
        \label{fig:thr-abl}
    }
    \hfill
    \subfigure[Left: ablation study of 1F1B scheduling. Right: memory footprint and throughput results of different 1F1B policies.]{
        \includegraphics[width=0.56\linewidth]{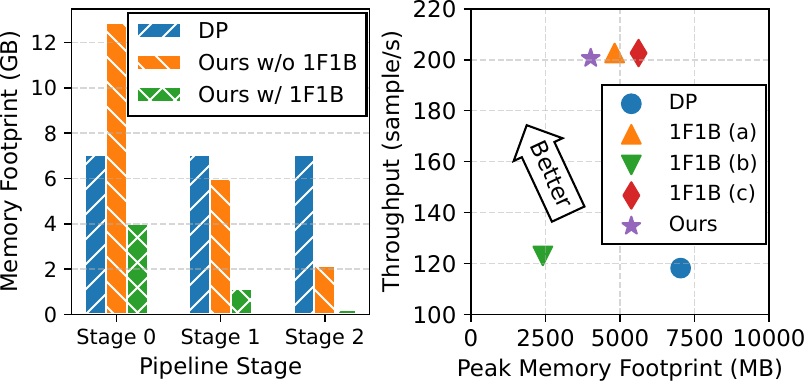}
        \label{fig:1f1b-eva}
    }
    \caption{Ablation study for each individual optimization technique in \S \ref{sec:execution}. $\times$ means out-of-memory error.}
\end{figure}

\textbf{Training Convergence Performance.}
We conducted training experiments on the EfficientNet-B1 and MobileNetV2 models using the CIFAR-10 image classification dataset in Env. B and C. We compared the time taken by Asteroid and baselines to achieve a target accuracy of 85\%. The results are demonstrated in Fig. \ref{fig:time2acc}.
Compared to synchronous methods, Asteroid achieves the target accuracy in a similar number of epochs as other baselines.
\revise{HetPipe's asynchronous updates induce parameter staleness, thereby impairing training accuracy. Consequently, more epochs are required to attain the target accuracy} \cite{wu2023hiflash, xie2019asynchronous}.
We observe that Asteroid achieves the target accuracy $1.2\times$-$6.1\times$ faster than all the baselines, attributed to either shorter per-epoch times or fewer required epochs.

\vspace{-0.3cm}
\subsection{Optimization Implication}
\label{sec:opt-impl}
This subsection investigates the performance boost of each individual optimization technique introduced in \S \ref{sec:execution}.

\textbf{Asteroid Parallelism Planning.} 
We conduct an ablation study using EfficientNet-B1 and MobileNetV2 on Env. C to assess the contributions of our inter-stage and intra-stage planning to the overall system performance, as depicted in Fig. \ref{fig:thr-abl}.
The naive approach without any planning optimization treated all devices as homogeneous and overlooked memory and bandwidth constraints. 
We observe that our inter-stage planning, which considers both computing heterogeneity between stages and gradient synchronization overhead, substantially boosts training throughput. Our intra-stage planning further enhances throughput by taking into account intra-stage heterogeneity and judiciously managing memory budgets to prevent run-time OOM issues.

\textbf{1F1B Micro-batch Scheduling.} 
Our analysis of the effectiveness of our 1F1B micro-batch scheduling revealed that, when applied to a 3-stage pipeline composed of three Jetson TX2 and used for training the EfficientNet-B1, our scheduling approach yields a much smaller per-stage memory footprint (estimated by Eq. (\ref{equ:mem})) compared to DP or conventional \textit{backward after forward} scheduling (Ours w/o 1F1B), as shown in Fig. \ref{fig:1f1b-eva}(Left).
To verify the superiority of our $K_p$ selection for stage $p$, we compare four policies: ($a$): $K_p=2\times (P-p)$. ($b$): $K_p=P-p$. ($c$): $K_p=2\times (P-p)+1$. (Ours): $K_p=2\times (P-p)-1$.
We can observe from Fig. \ref{fig:1f1b-eva}(Right) that our policy is sufficient to achieve a comparable training throughput as Policy $a$ and $c$, while having the smallest peak memory footprint.

\begin{figure}[t]
    \centering
    \begin{minipage}[b]{0.27\linewidth}
        \includegraphics[width=\linewidth]{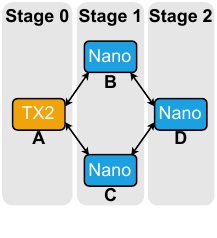}
    \end{minipage}
    \hfill
    \begin{minipage}[b]{0.67\linewidth}
        \includegraphics[width=\linewidth]{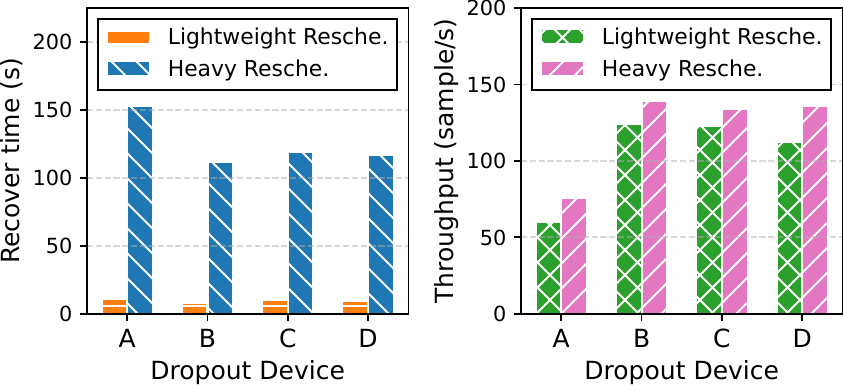}
        
    \end{minipage}
    \vspace{-10pt}
    \caption{Device grouping and performance of fault-tolerant module across diverse dropout scenarios.}
    \label{fig:eva-FT}
    \vspace{-10pt}
\end{figure}

\begin{figure}[t]
    \centering
    \setlength{\abovecaptionskip}{0.1cm}
    \setlength{\belowcaptionskip}{-0.4cm}
    \includegraphics[width=0.95\linewidth]{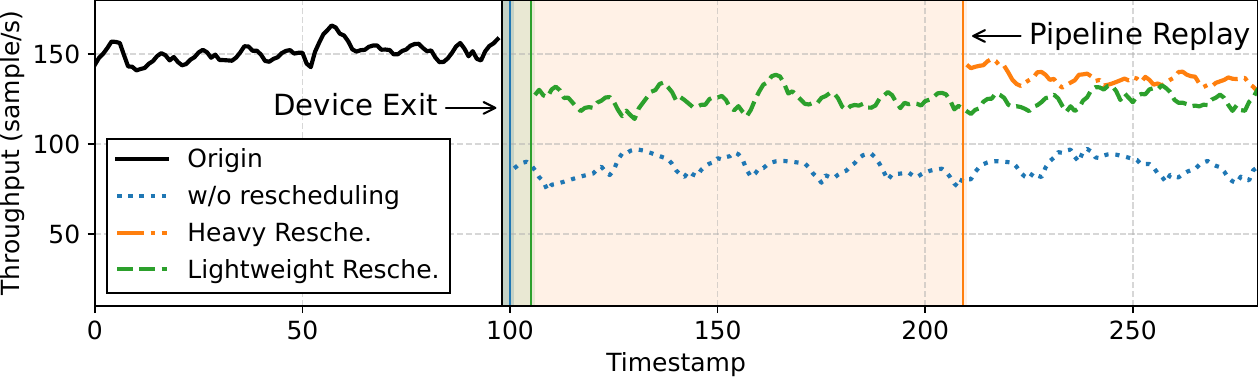}
    \caption{Throughput variation of different scheduling strategies when device B exits the training pipeline.}
    \label{fig:pipeline_replay}
    \vspace{-5pt}
\end{figure}

\vspace{-0.2cm}
\subsection{Fault Tolerance and Pipeine Replay}
\label{sec:fault}
We evaluate our lightweight fault-tolerant pipeline replay module with EfficientNet-B1 on Env. D, with device orchestration illustrated in Fig. \ref{fig:eva-FT}(Left).
We simulate the individual dropout of four devices, contrasting our lightweight approach with the heavy rescheduling.
\textit{Heavy rescheduling} involves aggregating stage models, rerunning the planning algorithm, and redistributing weights according to the new configuration. The planning algorithm is re-executed on the most powerful remaining device.
As illustrated in Fig. \ref{fig:eva-FT}(Mid) and \ref{fig:eva-FT}(Right), our mechanism recovers significantly faster than the heavy rescheduling and can achieve a comparable system throughput across diverse scenarios.
Specifically, Fig. \ref{fig:pipeline_replay} showcases the real-time training throughput over a time window, wherein we deliberately halt device B at the 100th timestamp. We observe that our lightweight mechanism can achieve $90\%$ throughput of heavy rescheduling, while recovering $14\times$ faster.

\vspace{-0.3cm}
\subsection{Scalability}
\begin{figure}[t]
    \centering
    \setlength{\abovecaptionskip}{0.1cm}
    \setlength{\belowcaptionskip}{-0.3cm}
    \includegraphics[width=0.95\linewidth]{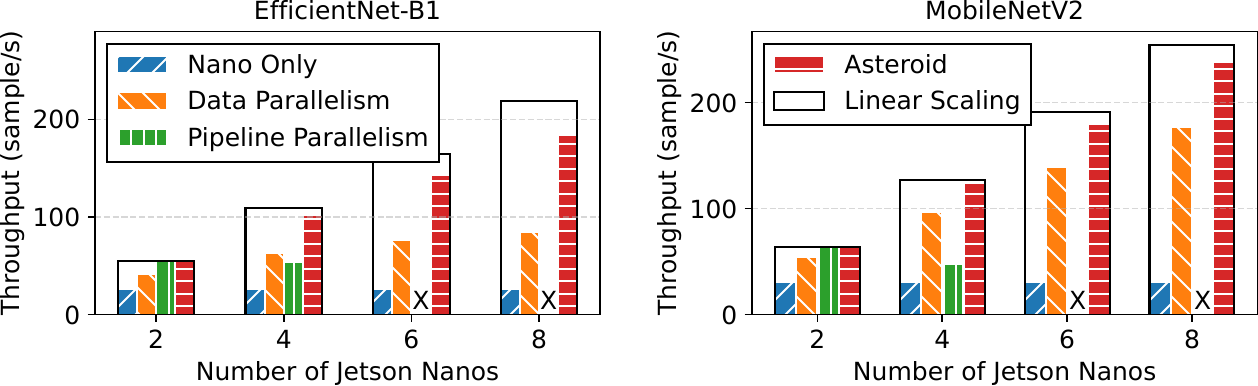}
    \caption{Throughput results with a varying number of Jetson Nanos. $\times$ means out-of-memory error.}
    \label{fig:scability}
    \vspace{-5pt}
\end{figure}

We analyze the scalability of Asteroid on an 8-node homogeneous Jetson Nano cluster with a fixed micro-batch size of $32$ per device (e.g. B=128 for 4 Jetson Nano) and conduct experiments on both EfficientNet-B1 and MobileNetV2 under 100Mbps network bandwidth. 
Fig. \ref{fig:scability} shows that Asteroid's hybrid parallelism exhibits substantial scalability even under a bandwidth-limited environment, attaining a throughput enhancement of $1.3\times$ - $2.2\times$ in comparison to DP when apply to EfficientNet-B1 and exhibiting near-linear scalability performance with MobileNetV2.
\revise{Gpipe's partitioning algorithm overlooks the sizes of intermediate tensors at partition points, making inter-stage communication the bottleneck in PP. Consequently, a 4-stage pipeline suffers from lower throughput than its 2-stage counterpart.
Moreover, despite employing the 1F1B memory optimization technique, OOM errors arise when scaling up to 6 devices.}

\begin{table}[t]
\small
\caption{Planning for Env. C across diverse models.}
\vspace{-10pt}
\begin{tabular}{ccccc}
\hline
Model & EffNet & MobileNet & ResNet & Bert \\ \hline
\begin{tabular}[c]{@{}c@{}}Planning time\end{tabular} & 480sec & 261sec & 192sec & 69sec \\ \hline
\end{tabular}
\label{tab:planning}
\vspace{-10pt}
\end{table}

\begin{table}[]
\small
\caption{Total profiling times of four models on devices.}
\vspace{-10pt}
\begin{tabular}{cccc}
\hline
Edge device & Jetson Nano & Jetson TX2 & Jetson NX \\ \hline
Profiling time & 82min & 51min & 25min \\ \hline
\end{tabular}
\label{tab:profiling}
\vspace{-10pt}
\end{table}

\vspace{-0.3cm}
\subsection{System Overhead}
\label{sec:complexity}
\textbf{Energy Consumption.}
We measure the energy consumption during the EfficientNet-B1 training process on Env. D. The experiment shows that Asteroid consumes $\sim$0.13 J per training sample and achieves $2$-fold reductions in energy consumption compared to DP. This improvement can be attributed to both the reduced on-device training time and network communication overhead.

\textbf{Planning Overhead.}
We evaluate the overhead of Asteroid's parallelism planning on Env. C involving six edge devices. The planning time across diverse models using Jetson NX is detailed in Table \ref{tab:planning}.
Specifically, \textit{Asteroid Planner} takes 480 seconds to partition the 213-layer EfficientNet-B1 optimally, compared to 69 seconds for the 56-layer Bert-small. As the number of model layers increases, the planning time rises significantly.
To mitigate this overhead in practical deployment, we can partition models at a coarser granularity (e.g., residual blocks), thereby narrowing the search space.

\textbf{Profiling Overhead.}
Profiling is another major overhead in Asteroid. 
We profile models including EfficientNet-B1, MobileNetV2 and Bert-small with batch sizes ranging from 1 to 256, as well as ResNet50 with batch sizes from 1 to 32. The total profiling times for all models on each edge devices are detailed in Table \ref{tab:profiling}.
The profiling overhead can be linearly scaled down by concurrent profiling on more devices.

Notably, both planning and profiling stages are one-shot offline processes and their outputs can be stored and reused. The overhead of these offline stages can be amortized across thousands of training iterations. 

\vspace{-0.3cm}
\section{Related Work}
\textbf{On-Device DNN Training.} 
POET \cite{patil2022poet} manages to fine-tune a BERT model on an embedded device in both training and energy efficiency. Lin et al. \cite{lin2022ondevice} make on-device training possible with only 256KB of memory. Sage and Melon \cite{gim2022memory, wang2022melon} adopt hybrid memory managing and saving techniques such as operator fusion and dedicated memory pool to address the memory constraints. 
Mandheling \cite{xu2022mandheling} adopt mixed-precision training with DSP offloading for learning acceleration.

\textbf{Collaborative Edge Computing for DNNs.} 
BlastNet, CoDL and $\mu\text{Layer}$ \cite{ling2022blastnet, kim2019mulayer, jia2022codl} perform a collaborative DL inference on CPU and GPU concurrently. CoEdge, DeepThings and MoDNN \cite{zeng2020coedge, zhao2018deepthings, mao2017modnn} distributed execution of CNN-based inference applications on resource-constrained edge clusters. 
EDDL \cite{hao2021eddl} adopts DP training across embedded devices.

\begin{figure}[t!]
    \centering
    \setlength{\abovecaptionskip}{0.1cm}
    \setlength{\belowcaptionskip}{-0.4cm}
    \includegraphics[width=0.9\linewidth]{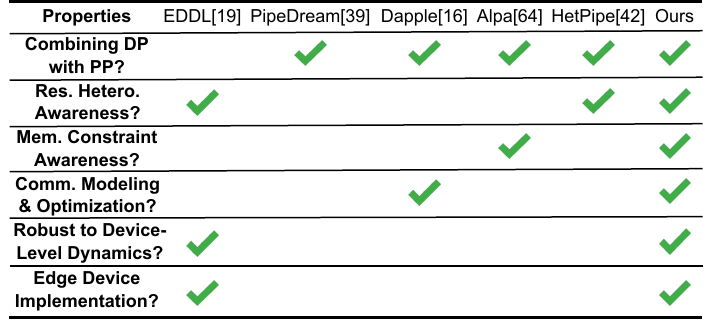}
    \caption{Comparing Asteroid with other systems.}
    \label{fig:related_work}
    \vspace{-0.2cm}
\end{figure}

\textbf{Parallel DNN Training in Datacenter.}
DP \cite{li2014communication, sergeev2018horovod, rajbhandari2020zero, goyal2017accurate} is the most extensively used distributed training method in datacenter.
PP \cite{huang2019gpipe, kim2020torchgpipe} has been proposed to conquer the memory issues of training large-scale transformer-based models.
\highlight{HPP and HDP \cite{narayanan2019pipedream, fan2021dapple, narayanan2021efficient,jia2022whale, luo2022efficient, team2020deepspeed, zheng2022alpa, park2020hetpipe} which combine merits of both DP and PP has been further proposed to address the resource efficiency and scalability.}
Fig. \ref{fig:related_work} provides a comparison of Asteroid with other existing works, emphasizing the distinctions between them.

\vspace{-0.3cm}
\section{Conclusion}
\label{sec:concolusion}
This paper proposes Asteroid for collaborative DNN training across heterogeneous and resource-constrained edge devices. Asteroid addresses multiple challenges faced in edge environments and achieves $12.2\times$ faster training than traditional methods and $2.1\times$ faster than state-of-the-art HPP methods. 


\balance
\bibliographystyle{ACM-Reference-Format}
\bibliography{reference}


\end{document}